\documentclass{aa}
\usepackage{txfonts}
\usepackage{graphicx}
\usepackage{natbib}

\bibpunct{(}{)}{;}{a}{}{,}

\begin{document}

\newcommand{\dd}{\mathrm{d}}
\newcommand{\eps}{\varepsilon}
\newcommand{\crp}{\mathrm{CRp}}
\newcommand{\cre}{\mathrm{CRe}}
\newcommand{\e}{\mathrm{e}}
\newcommand{\p}{\mathrm{p}}
\newcommand{\me}{m_\e}
\newcommand{\ag}{\alpha_\gamma}

\title{Constraining the population of cosmic ray protons in cooling \\
  flow clusters with $\gamma$-ray and radio observations: \\
  Are radio mini-halos of hadronic origin?}
\titlerunning{Constraining cosmic ray protons in cooling flow clusters}
\author{Christoph Pfrommer \and Torsten A. En{\ss}lin}
\authorrunning{C. Pfrommer \& T. A. En{\ss}lin}
\institute{Max-Planck-Institut f\"{u}r Astrophysik,
Karl-Schwarzschild-Str.1, Postfach 1317, 85741 Garching, 
Germany} 
\offprints{Christoph Pfrommer, \\
\email{pfrommer@mpa-garching.mpg.de}}
\date{Received 18 April 2003 / Accepted 12 September 2003}

\abstract{ We wish to constrain the cosmic-ray proton (CRp) population in
  galaxy clusters.  By hadronic interactions with the thermal gas of the
  intra-cluster medium (ICM), the CRp produce $\gamma$-rays for which we
  develop an analytic formalism to deduce their spectral distribution.
  Assuming the CRp-to-thermal energy density ratio $X_\crp$ and the CRp
  spectral index to be spatially constant, we derive an analytic relation
  between the $\gamma$-ray and bolometric X-ray fluxes, $\mathcal{F}_{\gamma}$
  and $F_\mathrm{X}$. Based on our relation, we compile a sample of suitable
  clusters which are promising candidates for future detection of $\gamma$-rays
  resulting from hadronic CRp interactions. Comparing to EGRET upper limits, we
  constrain the CRp population in the cooling flow clusters Perseus and Virgo
  to $X_\crp < 20\%$. Assuming a plausible value for the CRp diffusion
  coefficient $\kappa$, we find the central CRp injection luminosity of M~87 to
  be limited to $ 10^{43}\mbox{ erg s}^{-1} \, \kappa/(10^{29} \mbox{
    cm}^{2}\mbox{ s}^{-1})$.  The synchrotron emission from secondary electrons
  generated in CRp hadronic interactions allows even tighter limits to be
  placed on the CRp population using radio observations. We obtain excellent
  agreement between the observed and theoretical radio brightness profiles for
  Perseus, but not for Coma without a radially increasing CRp-to-thermal energy
  density profile. Since the CRp and magnetic energy densities necessary to
  reproduce the observed radio flux are very plausible, we propose synchrotron
  emission from secondary electrons as an attractive explanation of the radio
  mini-halos found in cooling flow clusters. This model can be tested with
  future sensitive $\gamma$-ray observations of the accompanying
  $\pi^0$-decays. We identify Perseus (A~426), Virgo, Ophiuchus, and Coma
  (A~1656) as the most promising candidate clusters for such observations.
}
\maketitle

\section{Introduction\label{sec:intro}}

Cooling flows are regions where the influence of non-thermal intra-cluster
medium (ICM) components such as magnetic fields and cosmic rays may be
strongest within a galaxy cluster owing to strong observed magnetic fields,
central active galaxies, and increasing non-thermal-to-thermal energy ratio due
to rapid thermal cooling processes.
They are also regions where such components are best detectable due to the high
gas density which allows for secondary particle production in hadronic
interactions of cosmic ray nuclei with the ambient gas. 
By the term cooling flow we do not rely on specific models but only on observed 
properties such as declining temperature gradients and enhanced electron density
profiles towards the center of the cluster. 

Non-thermal relativistic particle populations such as cosmic ray electrons
(CRe) and protons (CRp) can be injected into the ICM mainly by three different
processes \citep[following][]{2002BrunettiTaiwan} which produce radio
signatures that differ morphologically as well as spectrally: 
\begin{enumerate}
\item {\bf Shock acceleration:} 
Natural acceleration mechanisms providing relativistic particles are strong
structure formation and merger shocks \citep[e.g.,][]{1980A&AS...39..215H,
  1999ApJ...520..529S}.
Detailed studies have been undertaken on shocks of cosmological scales
\citep{2000ApJ...542..608M, 2000ApJ...535..586T}. Fermi~I
acceleration processes of CRe at these shock fronts produce large scale
extended peripheral radio relics as proposed by \citet{1998AA...332..395E}.
For instance two prominent relics in \object{Abell~3667}
\citep{1997MNRAS.290..577R} were successfully reproduced in a simulation of
this process by \citet{1999ApJ...518..603R}.
\item {\bf Reaccelerated electrons:} Secondly, reacceleration processes of
  mildly relativistic CRe ($\gamma\simeq 100-300$) being injected over
  cosmological timescales into the ICM by sources like radio galaxies,
  supernova remnants, merger shocks, and galactic winds can provide an
  efficient supply of highly-energetic CRe.  Owing to their long lifetimes of a
  few times $10^9$ years these mildly relativistic CRe can accumulate within
  the ICM \citep[see][ and references therein]{2002mpgc.book....1S}, until they
  experience continuous in-situ acceleration either via shock acceleration or
  resonant pitch angle scattering by turbulent Alv\'en waves as originally
  proposed by \citet{1977ApJ...212....1J}, reconsidered by
  \citet{1987A&A...182...21S}, and lately by \citet{2002ApJ...577..658O}.
  These acceleration processes of CRe possibly yield extended radio halos
  centered on the cluster \citep{2001MNRAS.320..365B} while there are also
  suggestions that radio mini-halos within a cooling flow cluster originate
  from these processes \citep{2002A&A...386..456G}.  There is also evidence
  that reacceleration processes acting on fossil radio plasma produces small
  filamentary radio relics at the cluster periphery, so-called revived radio
  ghosts \citep{2001A&A...366...26E, 2002MNRAS.331.1011E} presumably by
  adiabatic compression in shock waves.
\item {\bf Particles of hadronic origin:} Eventually, CRp can interact
  hadronically with the thermal ambient gas producing secondary electrons,
  neutrinos, and $\gamma$-rays in inelastic collisions taking place throughout
  the cluster volume which would generate radio halos through synchrotron
  emission \citep[first pointed out by][]{1980ApJ...239L..93D,
    1982AJ.....87.1266V}.  In the ICM the CRp have lifetimes of the order of
  the Hubble time \citep{1996SSRv...75..279V, 1997ApJ...477..560E,
    1997ApJ...487..529B}, long enough to diffuse away from the production site
  and to maintain a distribution over the cluster volume.  This process was
  reconsidered in more detail by \citet{1999APh....12..169B} and by
  \citet{2000A&A...362..151D}, the latter authors using numerical
  hydro-dynamical simulations including magnetic fields.  Recently,
  \citet{2001ApJ...559...59M} have performed cosmological simulations of
  cluster formation including injection processes of primary CRp.  These
  authors conclude that under certain conditions extended diffuse radio
  emission could be due to hadronically produced CRe. However, there are also
  claims that extended radio halos cannot be generated by secondary electrons
  due to the morphological steepness of predicted radio brightness profiles in
  contrast to observations \citep{2002BrunettiTaiwan}.  Besides constraining
  the population of CRp in the ICM, this work will present arguments for the
  hadronic origin of radio mini-halos or a substantial contribution of
  secondary electrons to these mini-halos. We further perform a parameter study
  which shows that the large cluster radio halos could be also of hadronic
  origin, provided the CRp-to-thermal energy density profile is radially
  increasing.
\end{enumerate}

It is very difficult to distinguish between contributions of these three
populations of cosmic ray (CR) particles to non-thermal particle populations,
especially if all of them account for injection of cosmic rays into the ICM in
different strength depending on underlying governing physical processes and
parameters. 
The hadronically produced CRe may be reaccelerated by shocks or cluster
turbulence and therefore mix up the different CRe populations.

Radio observations of the radio halo in the Coma cluster find a strong
steepening of the synchrotron spectrum with increasing radius
\citep{1993ApJ...406..399G}. This behavior is expected for a reaccelerated
population of CRe \citep{1999dtrp.conf..263B, 2001MNRAS.320..365B}. There is
also a report of radial spectral steepening in the case of the radio
mini-halo of Perseus according to \citet{Sijbring1993}.
This, however, could easily be an observational artifact owing to a poor
signal-to-noise ratio in the outer core parts of the cluster in combination
with the ambiguity of determining the large scale Fourier components owing to
the nonuniform coverage of the Fourier plane and missing short-baseline
information: the so-called ``missing zero spacing''-problem of interferometric
radio observations.  
By comparing the spectral index distribution of the three radio maps (92 cm, 49 
cm, and 21 cm), there seems to be likewise a possibility of radial spectral
flattening depending on the chosen radial direction. 
The hadronic electron model does not necessarily produce the radial
spectral steepening without fine-tuning.

{\bf Assumptions:}
The purpose of this work is to provide conceptually simple analytic instruments
for describing the spectral signatures in radio, X-rays, and $\gamma$-rays
resulting from inelastic cosmic ray ion collisions. It is especially important to
constrain the population of CRp within clusters of galaxies in order to
understand the governing physical processes of these objects and the important
theoretical implications for the non-thermal content of the ICM, i.e.~if
non-thermal CR pressure plays an important role in supporting the intra-cluster
ionized gas \citep{1997ApJ...477..560E}.
The assumptions of our models are:
\begin{itemize}
\item CRe are taken to originate from hadronic interactions of CRp with thermal
  ambient protons of the ICM and the CRp population is described by a power-law
  distribution in momentum. The origin of this population is not specified
  here, but CRp may be accelerated by shock waves of cluster mergers, accretion
  shocks \citep{1998APh.....9..227C}, or injected from radio galaxies into the
  ICM \citep{1984A&A...135..141V, 1997ApJ...477..560E, 1999APh....12..169B}, or
  result from supernova driven galactic winds \citep{1996SSRv...75..279V}.
\item In our isobaric model, the energy density of CRp is assumed to be
  proportional to the thermal energy density of the ICM.  In our scenario of
  adiabatic compression of CRp during the formation of the cooling flow this
  proportionality is imposed prior to the transition.  This assumption is
  reasonable if the thermal electron population and the CRp were energized by
  the same shock wave assuming that there is a constant fraction of energy
  going into the CRp population by such an acceleration process.  As a third
  model we take a single central point source injecting the CRp which results
  in a very peaked CRp profile \citep[compare][]{1999ApJ...525..603B,
    1999APh....12..169B}.
\item The CRp spectral index is assumed to be independent of position and
  therefore constant over the cluster volume. In some sense this represents an
  oversimplification which could be abandoned in order to reproduce some
  specific observational results, which however would be questionable without
  understanding the underlying physical processes.
\item The electron density and temperature profiles of the ICM are assumed to
  be spherically symmetric and were taken from the literature.  This assumption
  is justified in the case of $\gamma$-rays resulting from neutral pion decay
  because we use cluster volume averaged spectra in order to compare to
  observation, and is not severe in the case of radio emission, since the
  profiles are obtained from deprojected X-ray data.  The magnetic field
  configuration is assumed to be spherically symmetric on cluster core scales
  and follows the electron density with a power-law index as a free parameter
  within the suggested range \citep{2001A&A...378..777D, 1999A&A...348..351D}.
\item No reacceleration or diffusion process of CRe is taken into account in
  calculating the synchrotron and the inverse Compton (IC) emission. Therefore
  we provide conservative estimates for the flux.
\item The radio spectrum is taken to be quasi-stationary owing to the short
  electron cooling time which establishes a stationary CRe population on very
  short timescales. There is a one-to-one correspondence between the CRp power
  law index and that of the CRe population which is in addition determined by
  radiative synchrotron losses and IC cooling.
\end{itemize}

The paper is organized in two main parts: It starts with theoretic modeling of
$\gamma$-ray spectra resulting from hadronic CRp-p interactions and presents
models of synchrotron and IC radiation emitted by secondary electrons.  The
second part discusses the astrophysical application of this formalism to a
nearby cluster sample including cooling flow clusters.  After modeling the
spatial distribution of CRp within cooling flow clusters we constrain this
population by comparing to EGRET upper limits. We furthermore obtain limits on
the CRp population by the morphology of radio brightness profiles in the case
of Perseus and Coma.  Throughout this paper we assume the standard $\Lambda$CDM
cosmology with $\Omega_\mathrm{M} = 0.3$, $\Omega_\Lambda = 0.7$, and $H_0 =
70\, h_{70}\mbox{ km s}^{-1} \mbox{ Mpc}^{-1}$, where $h_{70}$ indicates the
scaling with $H_0$.

\section{Theoretic modeling of multi-frequency signatures resulting from
  hadronic CRp interactions} 

In order to study non-thermal emission from clusters we model the IC and
synchrotron radiation of secondary CRe produced in inelastic collisions by CRp
scattering off thermal nuclei as well as the $\gamma$-ray spectrum produced by
decaying pions being produced by these CRp-p collisions.  After introducing
  our definitions (Sect.~\ref{sec:def}), we develop an analytic formalism
  describing the decay of secondary neutral pions into two high-energy
  $\gamma$-rays (Sect.~\ref{sec: gamma-rays}).  Section~\ref{sec: fireball}
uses the analytical fireball model for inelastic CRp interactions with nuclei
of the intergalactic medium in the high-energy regime of CRp ($E_\p \gg
m_\p\,c^2$), following \citet{1994A&A...286..983M}.  Based on that we develop
in Sect.~\ref{sec: Dermer} an analytic formula describing the $\gamma$-ray
spectrum by parameterizing important effects near the pion threshold using an
approximate description developed by \citet{1986ApJ...307...47D,
  1986A&A...157..223D}, which combines isobaric \citep{1970Ap&SS...6..377S} and
scaling models \citep{1977PhRvD..15..820B,1981Ap&SS..76..213S} of the hadronic
reaction.   Using this formalism, an analytic
  $\mathcal{F}_{\gamma}$--$F_\mathrm{X}$ scaling relation is derived in the
  framework of a simple scenario of spatial distribution of CRp
  (Sect.~\ref{sec:F_g-F_X-scaling}). Finally, Sect.~\ref{sec: radio} deals
with radio and X-ray emission of secondary electrons being produced by decaying
charged pions.

\subsection{Definitions}
\label{sec:def}
Throughout the paper we use the following definitions for the differential
source function $q(\vec{r},E)$, the emissivity $j(\vec{r},E)$ and the volume
integrated quantities, respectively:

\noindent
\resizebox{\hsize}{!}{
\begin{minipage}[t]{0.5\hsize}
\begin{eqnarray*}
q(\vec{r},E) &=& \frac{\dd N}{\dd t\, \dd V\, \dd E}\,,\\
Q(E)         &=& \int \dd V\, q(\vec{r},E)\,,
\end{eqnarray*}
\end{minipage}\hfill
\begin{minipage}[t]{0.5\hsize}
\begin{eqnarray}
\rule{0mm}{5mm}   j(\vec{r},E) &=& E\, q(\vec{r},E)\,,\\
\rule{0mm}{6.2mm} J(E)         &=& E\, Q(E)\,,\label{Q}
\end{eqnarray}
\vspace{0.1cm}
\end{minipage}}
where $N$ denotes the integrated number of particles.
From the source function the integrated number density production rate of
particles $\lambda(\vec{r})$, the number of particles produced per unit time
interval within a certain volume, $\mathcal{L}$, and the particle flux
$\mathcal{F}$ can be derived. 
The definitions of the energy weighted quantities are denoted on the right hand
side, respectively,

\noindent
\resizebox{\hsize}{!}{
\begin{minipage}[t]{0.5\hsize}
\begin{eqnarray*}
\lambda(\vec{r})&=& \int\dd E\,q(\vec{r},E)\,,\\
\mathcal{L}     &=& \int\dd V\,\lambda(\vec{r})\,,\\
\mathcal{F}     &=& \frac{\mathcal{L}}{4\,\pi\,D^2}\,,
\end{eqnarray*}
\end{minipage}\hfill
\begin{minipage}[t]{0.5\hsize}
\begin{eqnarray}
\Lambda(\vec{r})&=& \int\dd E\,E\,q(\vec{r},E)\,,\\
L               &=& \int\dd V\,\Lambda(\vec{r})\,,\label{luminosity}\\
F               &=& \frac{L}{4\,\pi\,D^2}\,.\label{flux}
\end{eqnarray}
\vspace{0.1cm}
\end{minipage}}

\subsection{$\gamma$-ray spectrum from hadronic CRp interactions}
\label{sec: gamma-rays}

\subsubsection{Cosmic ray proton population}
\label{sec:CRp}

The differential number density distribution of a CRp population can be
described by a power-law in momentum $p_\p$,
\begin{equation}
\label{fp}
f_\p (\vec{r}, p_\p) \,\dd p_\p\,\dd V = 
\tilde{n}_\crp(\vec{r})\, \left(\frac{p_\p \,c}{\mbox{GeV}} \right)^{-\alpha_\p}\,
\left(\frac{c\,\dd p_\p}{\mbox{GeV}}\right)\, \dd V\,,
\end{equation}
where the tilde indicates that $\tilde{n}_\crp$ is not a real CRp number
density while it exhibits those dimensions.  We choose the normalization
$\tilde{n}_\crp(\vec{r})$ in such a way that the kinetic CRp energy density
$\eps_{\crp} (\vec{r})$ is proportional to the thermal energy density
$\eps_\mathrm{th} (\vec{r})$ of the ICM,
\begin{eqnarray}
\label{Xcrp scaling}
\eps_\crp(\vec{r}) &=&  X_\crp(\vec{r}) \, \eps_\mathrm{th}(\vec{r}) = 
\int_0^\infty \dd p \, f_\p(\vec{r}, p_\p)\,E_\mathrm{kin}(p_\p)\\ 
&=&   \frac{\tilde{n}_\crp(\vec{r})\,m_\p \,c^2}{2\,(\alpha_\p-1)}\,
\left(\frac{m_\p \,c^2}{\mbox{GeV}}\right)^{1-\alpha_\p}\mathcal{B}
\left(\frac{\alpha_\p-2}{2},\frac{3-\alpha_\p}{2}\right)\,.
\end{eqnarray}
The kinetic energy of CRp $E_\mathrm{kin}$ and the thermal energy density of
the ICM $\eps_\mathrm{th}$ are given by
\begin{eqnarray}
E_\mathrm{kin}(p_\p) &=&
\sqrt{p_\p^2\,c^2 + m_\p^2\,c^4} - m_\p\,c^2 ,\\
\eps_\mathrm{th} (\vec{r}) &=& \frac{3}{2}\,d_\e\,n_\e(\vec{r})\,k
T_\e(\vec{r})\,,\label{eps thermal}\\
\mbox{where}\quad
d_\e &=& 1 + \frac{1 - \frac{3}{4} X_\mathrm{He}}
{1 - \frac{1}{2} X_\mathrm{He}}
\end{eqnarray}
counts the number of particles per electron in the ICM using the primordial
$\element[][4]{He}$ mass fraction $X_\mathrm{He} = 0.24$, and
$\mathcal{B}(a,b)$ denotes the beta-function \citep{1965hmfw.book.....A}.  The
functional dependence of the CRp scaling parameter $X_\crp(\vec{r})$ is a
priori unknown. In order to draw astrophysical conclusions for the CRp
population in clusters of galaxies, we adopt three different models for the
spatial distribution of CRp later on in Sect.~\ref{sec: CRp models}.

In contrast to relativistic electrons which loose their energy on relatively short
time scales compared to the Hubble time through synchrotron emission in cluster
magnetic fields and IC scattering with photons of the microwave
background, the dominant energy loss mechanisms of CRp are electronic
excitations in the plasma \citep{1997ApJ...477..560E}, defining a cooling time
\citep{1972Physica....58..379G} 
\begin{eqnarray}
t_\mathrm{ee} &=& \left[ - \frac{1}{\gamma_\p} \left(
\frac{\dd \gamma_\p}{\dd t} \right)_\mathrm{ee} \right]^{-1} \\
&=&
\frac{m_\e \, c^3 \, m_\p \, \beta_\p\,\gamma_\p }{4 \, \pi\, e^4 \,n_\e}
\left[ \ln \left( \frac{2\gamma_\p m_\e c^2 \beta_\p^2}{\hbar
       \omega_\mathrm{pl}} \right) - \frac{\beta_\p ^2}{2} \right]^{-1},
\end{eqnarray}
where $\beta_\p c$ denotes the velocity of the proton, $\gamma_\p$ its
relativistic Lorentz factor, and $\omega_\mathrm{pl} = (4 \pi e^2 n_\e
/m_\e)^{1/2}$ the plasma frequency. 
Inserting typical values for cooling flows yields a lower
cutoff on the CRp momentum
\begin{equation}
\label{pmin}
p_\mathrm{min} = \beta_\p \gamma_\p \,m_\p\,c \simeq
0.2\,\left(\frac{t_\mathrm{age}}{\mbox{Gyr}} \right)
\left(\frac{n_\e}{10^{-2}\,\mbox{cm}^{-3}} \right)\, \mbox{GeV} c^{-1}.
\end{equation}
In general, this gives rise to a spatially dependent cutoff of the CRp momentum
which increases with time.

\begin{figure}[t]
\resizebox{\hsize}{!}{
\includegraphics{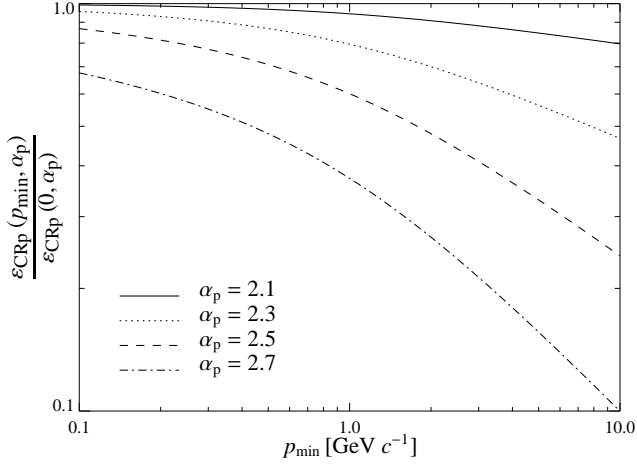}}
\caption{The ratio of CRp energy densities $\eps_\crp(p_\mathrm{min},\, \alpha_\p)$
  with and without a lower cutoff $p_\mathrm{min}$ in the CRp number density
  distribution function as a function of $p_\mathrm{min}$ for different values
  of the CRp spectral index $\alpha_\p$ (see eq.~(\ref{energyratio})). 
  For CRp the kinematically allowed threshold in order to produce
  $\pi^0$-mesons hadronically is given by $p_\mathrm{thr} = 0.78 \mbox{ GeV
  }c^{-1}$.}
\label{fig:energyratio}
\end{figure}

In order not to rely on too many assumptions, we do not impose a specific
momentum cutoff which is possible since the spectral index
$\alpha_\p$ varies in our model in between 2 and 3. 
Instead, we quantify the influence of a lower cutoff $p_\mathrm{min}$
on the population of CRp by taking the ratio of CRp energy densities
$\eps_\crp(p_\mathrm{min},\, \alpha_\p)$ with and without a lower
cutoff. This ratio as shown in Fig.~\ref{fig:energyratio} can be
written using the definition for the normalized lower CRp momentum
cutoff $\tilde{p} = \frac{p_\mathrm{min}}{m_\p\,c}$,
\begin{equation}
\label{energyratio}
\frac{\eps_\crp(\tilde{p},\, \alpha_\p)}{\eps_\crp(0, \alpha_\p)} =
\frac{\mathcal{B}_{x}\left(\frac{\alpha_\p-2}{2},
\frac{3-\alpha_\p}{2}\right) + 2\,\tilde{p}^{1-\alpha_\p} 
\left( \sqrt{1+\tilde{p}^2} - 1 \right)}{\mathcal{B} \left(\frac{\alpha_\p-2}{2},
\frac{3-\alpha_\p}{2}\right)}\,,
\end{equation}
where $\mathcal{B}_x(a,b)$ denotes the incomplete beta-function
\citep{1965hmfw.book.....A} with $x = (1+\tilde{p}^2)^{-1}$.  Combining
Fig.~\ref{fig:energyratio} and eq.~(\ref{pmin}) demonstrates the small
influence of Coulomb cooling to the CRp energy density within cooling flows.

\subsubsection{Fireball model}
\label{sec: fireball}

The CRp interact hadronically with the thermal background gas and produce 
pions with relative multiplicities $\xi_{\pi^0} = \frac{1}{2}\xi_{\pi^\pm}$
according to isospin symmetry and assuming thermal equilibrium of the pion
cloud in the center of mass \citep{1950PTP...5.570F}. The charged pions decay
into secondary electrons (and neutrinos) and the neutral pions into
$\gamma$-rays: 
\begin{eqnarray}
  \pi^\pm &\rightarrow& \mu^\pm + \nu_{\mu}/\bar{\nu}_{\mu} \rightarrow
  e^\pm + \nu_{e}/\bar{\nu}_{e} + \nu_{\mu} + \bar{\nu}_{\mu}\nonumber\\
  \pi^0 &\rightarrow& 2 \gamma \,.\nonumber
\end{eqnarray}
Only CRp above the kinematic threshold $p_\mathrm{thr} =
0.78 \mbox{ GeV }c^{-1}$ are able to produce pions hadronically and are
therefore visible through their decay products in both the $\gamma$-ray and
radio bands via radiative processes. 
Only the CRp population above this threshold is constrained by this work while
the lower energy part of this population in general can not be limited by 
only considering hadronic interactions.

In the high-energy limit for CRp ($E_\p \gg m_\p\,c^2$) the pion source
function resulting from hadronic CRp-p interactions can be calculated following 
\citet{1994A&A...286..983M} to be
\begin{eqnarray}
q_{\pi^0}(\vec{r},E_{\pi^0})\,\dd E_{\pi^0}\,\dd V &\approx& 
2^3\,\bar{\sigma}_\mathrm{pp}\,c\,n_\mathrm{N}(\vec{r})\, 
\frac{\tilde{n}_\crp(\vec{r})}{\mbox{GeV}} \nonumber \\
&&\times\, \left( \frac{6\,E_{\pi^0}}{\mbox{GeV}}\right)^{-\alpha_\gamma}\!
\dd E_{\pi^0}\,\dd V ,
\end{eqnarray}
where $\ag = 4/3\,(\alpha_\p - 1/2)$, $\bar{\sigma}_\mathrm{pp} = 32\,
\mbox{mbarn}$ is the inelastic p-p cross section, and $n_\mathrm{N}(\vec{r}) =
d_\mathrm{tar}\, n_\e(\vec{r}) = n_\e(\vec{r})/(1 - \frac{1}{2} X_\mathrm{He})$
is the target nucleon density in the ICM. 
The $\pi^0$-decay induced omnidirectional
(i.e. integrated over $4\,\pi$ solid angle) differential $\gamma$-ray source
function can be calculated in this energy regime assuming the decay products
are distributed isotropically in their rest frame, yielding
\begin{eqnarray}
q_\gamma(\vec{r},E_\gamma) &=& 2 \int_{E_\gamma + \frac{m_{\pi}^2 \,c^4}
{4\,E_\gamma}}^{\infty}\dd E_{\pi^0} \frac{q_{\pi^0}(\vec{r},E_{\pi^0})}
{\sqrt{E_{\pi^0}^2 - m_{\pi^0}^2 \,c^4}} \label{fireball}\\
&=& 2^3\, \bar{\sigma}_\mathrm{pp}\,c\,n_\mathrm{N}(\vec{r})\, 
\frac{\tilde{n}_\crp(\vec{r})}{\mbox{GeV}}
\, \left( \frac{6\,m_{\pi^0}\,c^2}{\mbox{GeV}}\right)^{-\alpha_\gamma}\!\!\!
\mathcal{B}_{x}\left(\frac{\ag}{2},\frac{1}{2}\right)\,, \nonumber\\
\mbox{where~~} x &=& \left(\frac{4\,E_\gamma\, m_{\pi^0} \,c^2}
{4\,E_\gamma^2 + m_{\pi^0}^2 \,c^4}\right)^2 .
\end{eqnarray}
Owing to Lorentz symmetry, this formula is valid for both
limiting energy regimes, $E_\gamma \gg m_{\pi^0} \,c^2/2$ and $E_\gamma \ll
m_{\pi^0}\,c^2/2$.
Because of an incomplete accounting of physical processes at the threshold of
pion production like the velocity distribution of CRp and momentum dependent
inelastic CRp-p cross section, eq.~(\ref{fireball}) overestimates the number of
$\gamma$-rays for energies around $E_\gamma \simeq m_{\pi^0}\,c^2/2$.

\subsubsection{Dermer's model}
\label{sec: Dermer}

In order to make detailed predictions for the $\pi^0$-decay induced $\gamma$-ray
spectrum, more realistic effects near the $\pi^0$-production threshold have to
be included. This was done by using the code COSMOCR originally designed for
cosmic ray studies by \citet{2001CoPhC.141...17M}. 
The underlying $\Delta_{3/2}$-isobaric model was shown to work well at low
proton energies \citep{1970Ap&SS...6..377S}. 
It assumes the CRp-p interaction to be mediated by the excitation of the
$\Delta_{3/2}$-resonance which subsequently decays into two protons and a
$\pi^0$-meson. The production spectrum of secondary $\pi^0$-mesons is given by
a convolution of the normalized $\Delta_{3/2}$-isobar mass spectrum represented
by a Breit-Wigner distribution with the energy distribution function.
The scaling model used at high energies \citep{1981Ap&SS..76..213S}
uses Lorentz invariant cross sections for charged and neutral pion production
in p-p interactions inferred from accelerator data. 
COSMOCR includes also the contribution of the two main kaon decay modes to
secondary pion spectra \citep[following][]{1998ApJ...493..694M} which are 
$K^\pm \to \mu^\pm + \nu_{\mu}/\bar{\nu}_{\mu}$ (63.5\%) and 
$K^\pm \to \pi^0 + \pi^\pm$ (21.2\%) where the latter channel also contributes
to the $\gamma$-ray source function.  

\begin{figure}[t]
\resizebox{\hsize}{!}{
\includegraphics{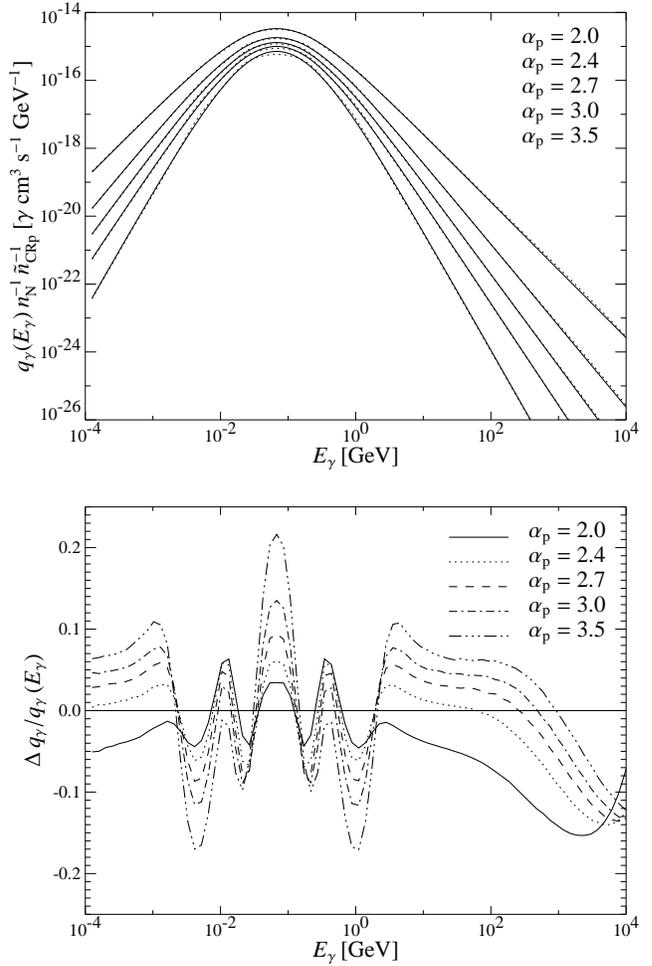}}
\caption{ {\bf a)} The omnidirectional (i.e. integrated over $4\,\pi$ solid
  angle) differential $\gamma$-ray source function $q_\gamma(E_\gamma)$
  normalized by the target number density $n_\mathrm{N}(\vec{r})$ and CRp
  normalization $\tilde{n}_\crp(\vec{r})$ in order to be independent of the spatial
  dependence of any specific model. The dotted lines show the simulated
  $\gamma$-ray spectra while the solid curves represent our models given by
  eq.~(\ref{q gamma}) with the spectral indices from top to bottom, $\alpha_\p
  \in \{2.0, ~2.4, ~2.7, ~3.0, ~3.5\}$.
  {\bf b)} Relative deviation of our analytic approach to simulated
  $\gamma$-ray spectra.}    
\label{fig:qfit}
\end{figure}

In order to derive an analytic formula describing the omnidirectional
differential $\gamma$-ray source function over the energy range shown in
Fig.~\ref{fig:qfit}, we keep the behavior of the spectrum in 
the fireball model for $E_\gamma \gg m_{\pi^0} \,c^2/2$ and
parameterize the detailed physics at the $\pi^0$-threshold by the shape
parameter $\delta_\gamma$ which smoothly joins the two power laws to the
asymptotic expansion of the $\mathcal{B}$-function of eq.~(\ref{fireball}),
yielding   
\begin{eqnarray}
\label{q gamma}
\lefteqn{
q_\gamma(\vec{r},E_\gamma)\,\dd E_\gamma\,\dd V\simeq
\sigma_\mathrm{pp}\,c\,n_\mathrm{N}(\vec{r})\,
\xi^{2-\alpha_\gamma}\,\frac{\tilde{n}_\crp(\vec{r})}{\mbox{GeV}}}  \\
  & & \times\,\frac{4}{3\,\ag}\,\left( \frac{m_{\pi^0}\,c^2}
      {\mbox{GeV}}\right)^{-\alpha_\gamma}
      \left[\left(\frac{2\, E_\gamma}{m_{\pi^0}\, c^2}\right)^{\delta_\gamma} +
      \left(\frac{2\, E_\gamma}{m_{\pi^0}\, c^2}\right)^{-\delta_\gamma}
      \right]^{-\ag/\delta_\gamma}\!\dd E_\gamma\,\dd V. \nonumber
\end{eqnarray}
The scaling behavior in the high-energy limit of Dermer's model can be
described by a constant pion multiplicity $\xi=2$ characterizing the two
leading pion jets leaving the interaction site in direction of the incident
protons diametrically and carrying the 
high longitudinal momenta owing to Lorentz contraction of the interacting
nuclei in the center of mass system and Heisenberg's uncertainty relation
\citep{1990epcp.book.....N}.
This assumption of constant pion multiplicity of the scaling model is
in contrast to the fireball model \citep{1994A&A...286..983M}, which assumes 
a state of hot quark-gluon plasma in thermal equilibrium after the hadronic
interaction subsequently ablating pions with multiplicities 
$\xi_{\pi^0} \simeq [(E_\p - E_\mathrm{th})/\mbox{GeV}]^{1/4}$, where
$E_\mathrm{th} = 1.22 \,\mbox{GeV}$ denotes the threshold energy for pion
production.

The $\gamma$-ray source function peaks at the energy of $m_{\pi^0}\, c^2/2
\simeq 67.5 \,\mbox{MeV}$.
It is well known, that the asymptotic slope of the $\gamma$-ray spectrum, 
characterized by its spectral index $\ag$, reproduces the spectral index of the
population of CRp, $\ag = \alpha_\p$ \citep{1986A&A...157..223D}. 
This is again in contrast to the fireball model which predicts a steeper
asymptotic slope in the $\gamma$-ray spectrum for $\alpha_\p > 2$, amounting
to $\ag = 4/3\,(\alpha_\p - 1/2)$.
In the following we restrict ourselves to Dermer's model because it is better
motivated by accelerator data.

By comparing the logarithm of the $\gamma$-ray source function of
eq.~(\ref{q gamma}) to numerically calculated spectra using COSMOCR we
recognized that the influence of the detailed 
physics at the threshold together with the kaon contribution can be modeled in
our semi-analytic approach in eq.~(\ref{q gamma}) by self-consistent scaling
relations for the shape parameter $\delta_\gamma$ and the effective inelastic
p-p cross section $\sigma_\mathrm{pp}$ including the kaon decay modes. 
The shape parameter $\delta_\gamma$ scales with the spectral index of the
$\gamma$-ray spectrum as  
\begin{equation}
\label{delta}
\delta_\gamma = 0.14 \,\alpha_\gamma^{-1.6} + 0.44\,,
\end{equation}
which models the functional behavior of the spectrum (compare
Fig.~\ref{fig:qfit}). 
The effective cross section $\sigma_\mathrm{pp}$ also depends on
$\alpha_\gamma$ which can be modeled by 
\begin{equation}
\label{sigmapp}
\sigma_\mathrm{pp} = 32 \times
\left(0.96 + \mathrm{e}^{4.4 \,-\, 2.4\,\alpha_\gamma}\right)\mbox{ mbarn}\,. 
\end{equation}
On the one hand, the enhanced contribution to the normalization of the
$\gamma$-ray source function $q_\gamma(E_\gamma)$ for flat spectral indices
$\alpha_\gamma$ is due to the larger contribution of the channel $p+p \to K^\pm
+ X$ relative to  $p+p \to \pi^\pm + X$ for larger energies, approaching
asymptotically a value of 27~\% at energies larger than 1 TeV
\citep{2001CoPhC.141...17M} which we did not account for a priori in our simple
model. 
Secondly, this scaling behavior also includes higher order contributions to the 
effective pion multiplicity for harder spectra characterized by a lower
spectral index $\alpha_\gamma$.

The effective description of the spectrum with the smooth peak characterized by
the shape parameter $\delta_\gamma$ starts to fail for very steep
spectra of $\alpha_\gamma > 3.5$ where relativistic kinematics at the
threshold plays a crucial role. Then the higher number of decaying low
energetic $\pi^0$-mesons results in a more concentrated peak on top of the
boosted broader distribution of decaying highly-energetic pions. 
The lower panel of Fig.~\ref{fig:qfit} shows that the relative deviation
of the semi-analytic approach of eq.~(\ref{q gamma}) to the simulated
$\gamma$-ray spectra amounts below 0.2 for the spectral range shown in
Fig.~\ref{fig:qfit}, which is sufficient for the purpose of our work. 

\subsection{Energy band integrated $\gamma$-ray luminosity: Analytic 
$\mathcal{F}_{\gamma}$--$F_\mathrm{X}$ scaling relation}
\label{sec:F_g-F_X-scaling}

In the following, we derive an analytic $\mathcal{F}_{\gamma}$--$F_\mathrm{X}$
scaling relation which should serve as an approximate estimate for a given
cluster of galaxies. As a simple scenario we choose the CRp energy density to
be a constant fraction of the thermal energy density, $\eps_\crp(\vec{r}) =
X_\crp \, \eps_\mathrm{th}(\vec{r})$.  However, this is not a fundamental
constraint for this scaling relation. Any other spatial dependency for the CRp
scaling parameter $X_\crp$ may be substituted instead of the assumed one.

The bolometric X-ray emission of the hot thermal intra cluster electrons is
given by the cooling function for thermal bremsstrahlung
\citep{1979rpa..book.....R},
\begin{eqnarray}
\label{Xray}
\Lambda_\mathrm{X}[n_\e(\vec{r}),T_\e(\vec{r})] 
  &=& \Lambda_0\, n_\e(\vec{r})^2 \,\sqrt{k T_\e(\vec{r})}\,, \\
\mbox{with~~}\Lambda_0 
  &=& \left(\frac{2\,\pi}{3\,m_\e}\right)^{1/2}
      \frac{2^5 \pi\,e^6 d_\mathrm{tar}}{3\,h\,m_\e\,c^3}\,
      Z^2\, \bar{g}_\mathrm{B}(T_\e) \label{Lambda0} \\
  &\simeq& 6.62 \times 10^{-24} \mbox{ erg s}^{-1}\mbox{ cm}^3\mbox{ keV}^{-1/2},
\end{eqnarray}
where $n_\e$ is the electron number density, $T_\e$ the temperature,
$d_\mathrm{tar}$ is the nucleon density in the ICM relative to the electrons
for primordial element composition, $Z$ the charge number\footnote{Setting $Z^2
  = 1$ in eq.~(\ref{Lambda0}) is correct for a plasma of primordial element
  composition which consists of hydrogen and helium only, because $\langle
  n_\mathrm{N} Z^2 \rangle = n_\mathrm{N}$ in this case. This is a reasonable
  approximation owing to the small contamination of heavier elements in the
  ICM.}  and $\bar{g}_\mathrm{B}\simeq 1.2$ is the frequency and velocity
averaged Gaunt factor.

In order to obtain the integrated $\gamma$-ray source density $\lambda_\gamma$
for pion decay induced $\gamma$-rays the $\gamma$-ray source function
$q_\gamma(\vec{r}, E_\gamma)$ in eq.~(\ref{q gamma}) can be integrated over an
energy interval yielding 
\begin{eqnarray}
\label{lambda_gamma}
\lambda_\gamma(\vec{r}, E_1, E_2) &=& \int_{E_1}^{E_2} \dd E_\gamma\,
q_\gamma(\vec{r}, E_\gamma) \\
&=& A_\gamma(\alpha_\p)\, N_\gamma(\alpha_\p)\, 
X_\crp\,n_\e^2(\vec{r})\,k T_\e(\vec{r})\,,\\
\mbox{where~~} 
A_\gamma(\alpha_\p) &=& \frac{\sigma_\mathrm{pp}(\alpha_\p)\,
      d_\e\,d_\mathrm{tar}\,m_{\pi^0}\,c}
      {\mbox{GeV }m_\p}\,,\label{Agamma}\\
N_\gamma(\alpha_\p) &=&
\frac{\left(\frac{m_{\pi^0}\,c^2}{\mbox{GeV}}\right)^{-\alpha_\gamma}
      \left[\mathcal{B}_x\left(\frac{\alpha_\gamma + 1}{2\,\delta_\gamma},
      \frac{\alpha_\gamma - 1}{2\,\delta_\gamma}\right)\right]_{x_1}^{x_2}
      (\alpha_\p-1)}
     {\left(\frac{m_\p\,c^2}{\mbox{GeV}}\right)^{1-\alpha_\p}
      \mathcal{B}\left(\frac{\alpha_\p-2}{2},
      \frac{3-\alpha_\p}{2}\right)\,2^{\alpha_\gamma -2}\, 
      \alpha_\gamma\,\delta_\gamma}\,,\label{Ngamma}\\
\mbox{and~~}
x_i &=& \left[1+\left(\frac{m_{\pi^0}\,c^2}{2\,E_i}
      \right)^{2\,\delta_\gamma}\right]^{-1} \mbox{~for~~}
      i \in \{1,2\}\,.
\end{eqnarray}
Here we introduced the abbreviation 
\begin{equation}
[f(x)]_{x_1}^{x_2} = f(x_2) - f(x_1)\,.
\end{equation}
Assuming Dermer's model the $\gamma$-ray spectral index scales as 
$\alpha_\gamma = \alpha_\p$ in contrast to the fireball model where 
$\ag = 4/3\,(\alpha_\p - 1/2)$.  
The shape parameter $\delta_\gamma$ is given by the
$\alpha_\gamma$--$\delta_\gamma$ scaling relation in eq.~(\ref{delta})
which strictly holds for Dermer's model, but should also be valid for the
extended fireball model.

Comparing the integrated $\gamma$-ray source density $\lambda_\gamma(E_1, E_2)$
of eq.~(\ref{lambda_gamma}) to that of thermal bremsstrahlung
(eq.~(\ref{Xray})) we obtain an analytic
$\mathcal{F}_{\gamma}$--$F_\mathrm{X}$ scaling relation for the ratio
of $\gamma$-ray fluxes $\mathcal{F}_{\gamma}$ and bolometric X-ray fluxes
$F_\mathrm{X}$,
\begin{equation}
\label{scaling}
\frac{\mathcal{F}_{\gamma}(E_1<E<E_2)}
     {F_\mathrm{X}^\mathrm{bol} \mbox{ erg}^{-1}} =
    \frac{A_\gamma(\alpha_\p)\, N_\gamma(\alpha_\p)}
     {\Lambda_0 \mbox{ keV}^{-1/2}\mbox{ erg}^{-1}}
  \,\left(\frac{\langle k T_\e\rangle}{\mbox{keV}}\right)^{1/2} X_\crp\,,
\end{equation}
where the prefactor is appropriately scaled yielding a dimensionless number which
consists of $A_\gamma(\alpha_\p)$ (eq.~(\ref{Agamma})), 
$N_\gamma(\alpha_\p)$ (eq.~(\ref{Ngamma})), and $\Lambda_0$ (eq.~(\ref{Lambda0})).
The $\mathcal{F}_{\gamma}$--$F_\mathrm{X}$ ratio scales linearly with
the scaling parameter $X_\crp$ given by eq.~(\ref{Xcrp scaling}) and is
independent of the underlying cosmology, however not of redshift due to the
K-correction. 
Inferred values for the expected $\gamma$-ray flux $\mathcal{F}_{\gamma}$ are
consistent with those obtained by \citet{1997ApJ...477..560E} for the spectral
index of our Galaxy $\alpha_\gamma = 2.7$.

\subsection{Stationary spectrum of hadronically originating secondary
  electrons} 
\label{sec: radio}

This section is based on a formalism developed in \citet{2000A&A...362..151D}.
The steady-state $\cre$ spectrum is governed by injection of secondaries and
cooling processes so that it can be described by the continuity equation
\begin{equation}
\frac{\partial }{\partial E_\e} \left( \dot{E_\e}(\vec{r},E_\e) f_\e
(\vec{r},E_\e) \right) = q_\e(\vec{r}, E_\e)\,.
\end{equation}
For $\dot{E_\e}(\vec{r},p) < 0$ this equation is solved by
\begin{equation}
f_\e (\vec{r},E_\e) = \frac{1}{|\dot{E_\e}(\vec{r},E_\e)|} \int_{E_\e}^\infty
\! dE_\e'  q_\e(\vec{r}, E_\e')\,.
\end{equation}
The cooling of the radio emitting $\cre$ is dominated by synchrotron and
inverse Compton losses giving
\begin{equation}
- \dot{E_\e}(\vec{r},E_\e) = \frac{4\,\sigma_\mathrm{T}\, c}{3\,m_\e^2\,c^4} \left(
\frac{B^2 (\vec{r})}{8\,\pi} + \frac{B^2_\mathrm{CMB}}{8\,\pi}
\right)\,E_\e^2\,,   
\end{equation}
where $\sigma_\mathrm{T}$ is the Thomson cross section, $B(\vec{r})$
is the local magnetic field strength and $B^2_{\rm CMB}/(8\,\pi)$ is the energy
density of the cosmic microwave background expressed by an equivalent field
strength $B_{\rm CMB} = 3.24\, (1+z)^2\mu\mbox{G}$.
The $\cre$ population above a GeV is therefore described by a power-law
spectrum 
\begin{eqnarray}
\label{fe}
f_\e (\vec{r},E_\e) &=& \frac{\tilde{n}_\cre(\vec{r})}{\rm GeV}
\,\left( \frac{E_\e}{\rm GeV} \right)^{-\alpha_\e}\,, \\
\tilde{n}_\cre(\vec{r}) &=& \frac{2^7 \,\pi\, 16^{-(\alpha_\e-1)}
\,A_\mathrm{mod}}{\alpha_\e - 2}\,
\frac{\sigma_\mathrm{pp}\, m_\e^2\,c^4}
     {\sigma_\mathrm{T} \,{\rm GeV}} \,
\frac{n_\mathrm{N}(\vec{r})\, \tilde{n}_\crp(\vec{r})}
     {B^2(\vec{r}) + B^2_{\rm CMB}}\,,\label{Ne} \\
\alpha_\e &=& \left\{\begin{array}{l@{\quad\mbox{in }}l}
            \alpha_\p + 1 & \mbox{Dermer's model}\,,\\
            \frac{4}{3}\,\alpha_\p + \frac{1}{3} & \mbox{the fireball model}\,,
            \end{array}\right.\\
A_\mathrm{mod}&=& \left\{\begin{array}{l@{\quad\mbox{in }}l}
                1 & \mbox{Dermer's model}\,,\\
                3\,\left(\frac{3}{2}\right)^{-(\alpha_\e-1)} &
                \mbox{the fireball model}\,.
                \end{array}\right.
\end{eqnarray}
For the sake of consistency, we use Dermer's model throughout the paper where
the effective cross section $\sigma_\mathrm{pp}$ is given by
eq.~(\ref{sigmapp}) in contrast to the fireball model where $\sigma_\mathrm{pp}
= 32$~mbarn. The approach of the scaling relation of eq.~(\ref{sigmapp})
is approximately valid for CRe although the decay channels
of charged kaons provide a stronger contribution to the
$\pi^\pm$ branching ratio relative to $\pi^0$-mesons
resulting also in slightly higher injection rates for electrons and positrons. 
Differences in normalization and radio brightness morphology due to the
different models governing the CRp-p interaction are small and irrelevant for
our conclusions.

\subsubsection{Synchrotron emission of secondary electrons}

The synchrotron emissivity $j_\nu$ at frequency $\nu$ and per steradian of a
power law distribution of CRe (eq.~(\ref{fe})) in an isotropic distribution of
magnetic fields and electrons within the halo volume 
\citep[eq.~(6.36) in][]{1979rpa..book.....R}, is obtained after averaging over
an isotropic distribution of electron pitch angles yielding
\begin{equation}
\label{jnu}
   j_\nu(\vec{r})=c_2(\alpha_\e)\, \tilde{n}_\cre(\vec{r}) \, 
   B(\vec{r})^{\alpha_\nu+1}
   \left(\frac{\nu}{c_1}\right)^{-\alpha_\nu}
\end{equation}
with $c_1=3\,e\,{\mbox{GeV}^2}/(2\,\pi\, m_\e^3\,c^5)\,$,
\begin{equation}
   c_2(\alpha_\e)=\frac{\sqrt{3\,\pi}}{32\,\pi}\frac{e^3}{m_\mathrm{e}c^2}
               \frac{\alpha_\e+\frac{7}{3}}{\alpha_\e+1}
   \frac{\Gamma\left( \frac{3\alpha_\e-1}{12}\right)
         \Gamma\left( \frac{3\alpha_\e+7}{12}\right)
         \Gamma\left( \frac{\alpha_\e+5}{4}\right)}
        {\Gamma\left( \frac{\alpha_\e+7}{4}\right)},
\end{equation}
where $\Gamma(a)$ denotes the $\Gamma$-function \citep{1965hmfw.book.....A}
and $\alpha_\nu = (\alpha_\e-1)/2 = \alpha_\p/2$ in Dermer's model.
In our models the magnetic field $B(r)$ was assumed to be spherically symmetric
on cluster core scales and and to follow the electron density $n_\e(r)$ 
\citep{1999A&A...348..351D, 2001A&A...378..777D}:
\begin{equation}
\label{magnetic field}
B(r) = B_0 \,\left[\frac{n_\e(r)}{n_\e(0)}\right]^{\alpha_B},
\end{equation}
where $B_0$ and $\alpha_B$ are free parameters in our model.
Assuming the radio emissivity $j_\nu (\vec{r})$ in eq.~(\ref{jnu}) to be only a
function of radius, then the line of sight integration yields the surface
brightness of the radio halo 
\begin{equation}
\label{Snu}
S_\nu(r_\bot) = 2 \int_{r_\bot}^R \frac{j_\nu(r)\,r\,\dd r}
{\sqrt{r^2 - r_\bot^2}}\,.
\end{equation}

\subsubsection{Inverse Compton emission of secondary electrons}

\begin{figure}[t]
\resizebox{\hsize}{!}{
\includegraphics{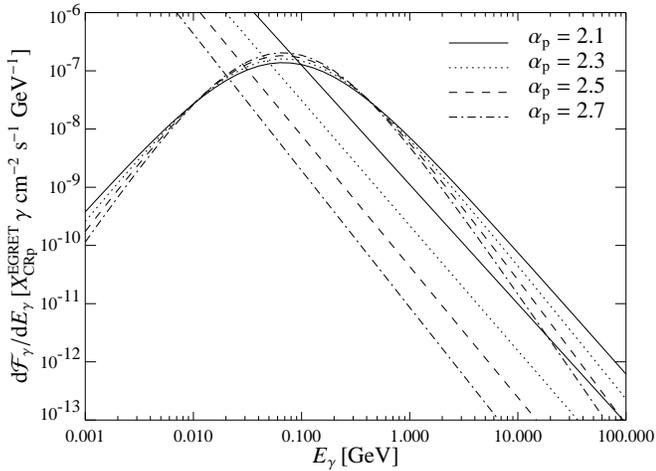}}
\caption{The simulated differential flux of $\gamma$-rays from Perseus reaching
  the Earth. Shown are upper limits of the IC emission of secondary CRe
  (power-laws, assuming zero magnetic field) as well as
  pion decay induced $\gamma$-ray emission (represented by broad distribution
  centered on $E_\mathrm{peak}\simeq 67.5 \mbox{ MeV}$).
  The normalization of the spectra differing in their values of the CRp spectral
  index $\alpha_\gamma=\alpha_\p$ (Dermer's model) depends on the assumed
  scaling between CRp and thermal energy density. 
  We fix this scaling parameter $X_\crp$ assuming the isobaric model
  (Sect.~\ref{sec: isobaric}) by comparing the integrated flux above
  100 MeV to EGRET upper limits \citep[see][]{2003ApJ...588..155R}.} 
\label{fig:qgamma}
\end{figure}

The source function $q_\mathrm{IC}$ owing to IC scattering of 
cosmic microwave background (CMB) photons off an isotropic power law
distribution of hadronically originating CRe (eq.~(\ref{fe})) is 
\citep[derived from eq.~(7.31) in][ in the case of Thomson scattering]
{1979rpa..book.....R},
\begin{eqnarray}
\label{IC}
q_\mathrm{IC}(\vec{r},E_\gamma) &=& \tilde{q}(\vec{r})\,
f_\mathrm{IC} (\alpha_e)\, 
\left(\frac{m_\e\, c^2}{\mbox{GeV}}\right)^{1 - \alpha_\e}
\left(\frac{E_\gamma}{k T_\mathrm{CMB}}\right)^{-(\alpha_\nu+1)}, \\
f_\mathrm{IC} (\alpha_e) &=& \frac{2^{\alpha_\e+3}\, 
  (\alpha_\e^2 + 4\, \alpha_\e + 11)}
  {(\alpha_\e + 3)^2\, (\alpha_e + 5)\, (\alpha_e + 1)} \\
&&\times\,\Gamma\left(\frac{\alpha_e + 5}{2}\right)\,
  \zeta\left(\frac{\alpha_e + 5}{2}\right)\,,\\
\mbox{and }\tilde{q}(\vec{r}) &=& 
\frac{8\, \pi^2\, r_\e^2\, \tilde{n}_\cre(\vec{r})\,
      \left(k T_\mathrm{CMB}\right)^2\,}{h^3\, c^2}\,, 
\end{eqnarray}
where $\alpha_\nu = (\alpha_\e-1)/2$ denotes the spectral index, $r_\e = e^2
/(m_\e\, c^2)$ the classical electron radius, $\zeta(a)$ the Riemann
$\zeta$-function \citep{1965hmfw.book.....A}, and $\tilde{n}_\cre(\vec{r})$ is
given by eq.~(\ref{Ne}). After integrating over the IC emitting volume in the
cluster we obtain the particle flux $\mathcal{F}(E_\gamma)$ (see
eqs.~(\ref{luminosity}) and (\ref{flux})).  The same CRe population seen in the
radio band via synchrotron emission scatter CMB photons into the hard X-ray
regime. In the $\gamma$-ray spectrum, there is a point of equal contribution of
the IC spectrum of the CRe showing a decreasing slope of $-\alpha_\nu-1=
-\alpha_\p/2-1$ (assuming Dermer's model) and the pion decay induced
$\gamma$-ray spectrum being characterized by the rising slope $\alpha_\gamma =
\alpha_\p$ (see eq.~(\ref{q gamma})). In the high energy limit ($E_\gamma \gg
m_{\pi^0} \,c^2/2$), the pion decay induced $\gamma$-ray spectrum declines with
a slope of $-\alpha_\gamma= -\alpha_\p$ which is the same as the IC emission
for $\alpha_\p=2$ and slightly steeper for larger values of $\alpha_\p$ (for
illustration, see Fig.~\ref{fig:qgamma}).

\subsection{Summary and outline}
In the previous sections we developed an analytic description to compute the
neutral pion decay induced $\gamma$-ray spectrum from a CRp population over a
broad range of $\gamma$-ray energies extending from below MeV up to TeV.
Moreover, we presented a formalism of calculating the synchrotron and IC
emission from a stationary population of CRe resulting from hadronic CRp
interactions of the thermal gas of the ICM.  Assuming a constant scaling
  between kinetic CRp energy density and thermal energy density of the ICM we
derived an analytic $\mathcal{F}_{\gamma}$--$F_\mathrm{X}$ scaling relation for
the ratio of $\gamma$-ray flux to bolometric X-ray flux to obtain
observationally promising cluster candidates for constraining the CRp
population.  In order to obtain reliable flux estimates we are going to
introduce in the following three possible spatial distributions of the CRp,
whose population is either in fractional pressure equilibrium with the thermal
particle population (as assumed for selecting the clusters), experienced
adiabatic compression during the formation of the cooling flow cluster or is
shaped by diffusion away from a central source of CRp.  By modeling the
$\gamma$-ray emission of these particular clusters and comparing to EGRET upper
limits we are going to present bounds on the CRp population.  Furthermore, we
will derive upper bounds on the CRp population by radio synchrotron emission of
hadronically originating CRe and will compare azimuthally averaged radio
brightness profiles of the the Perseus radio-mini halo and the radio halo of
Coma.

\section{Astrophysical application to nearby clusters of galaxies using
  multi-frequency observations}

\subsection{The expected spectral index $\alpha_\p$\label{sec:consrtalpha}}

The spectral index of the ICM CRp population $\alpha_\p$ is not well
constrained by observations. However, because galaxy clusters are able
to store CRp for cosmological times \citep{1996SSRv...75..279V,
1997ApJ...477..560E, 1997ApJ...487..529B} the spectral index of the
global CRp population (allowing for spatial differentiation) is
expected to be that of the injection process, if no re-acceleration
processes modified the spectrum after injection. We discuss briefly
different possible CRp sources in galaxy clusters:\\
{\bf Structure formation shock waves} have generated most of the
thermal energy content of galaxy clusters. Therefore, it is plausible
to assume that they also produced most of the CR energy of clusters.
Shock acceleration is able to produce momentum power-law particle
distributions characterized by a spectral index, which is in the
test-particle picture of non-relativistic shock acceleration
\begin{equation}
\alpha_{\rm inj} = \frac{R+2}{R-1}\,,
\end{equation}
where $R \le 4$ is the shock compression factor. The lowest spectral
indices are therefore generated by the strongest shocks, which are
preferentially found in peripheral regions of the clusters
\citep{1998ApJ...502..518Q,2000ApJ...542..608M}. Thus, harder CRp
populations ($\alpha_{\rm inj} = 2.0 ...2.5$) are mostly injected into
the outskirts of clusters. However, motion of the ICM gas
transports them efficiently into the cluster centers
\citep{2001ApJ...559...59M}.\\
{\bf Injection by radio galaxies:} Active galactic nuclei (AGN) are
able to produce large amounts of relativistic plasma. The composition
of this plasma is not known, however, the presence of CRp is assumed
in and supported by many papers. The energetics of AGN is sufficient
to inject a significant CRp population into the thermal ICM
\citep{1997ApJ...477..560E, 1998AA...333L..47E, 1998APh.....9..227C,
1999ApJ...525..603B, 2000MNRAS.318..889W}
provided CRp are present in the radio plasma and are able to leave it
on cosmological short timescales. If the radio plasma releases all its
CRp, a moderately flat injection spectrum can be expected (say
$\alpha_{\rm inj} \approx 2.5$) since radio emission from radio
galaxies indicates flat CRe spectra. 
If, however, only a small fraction of the
CRp is able to leave the radio plasma diffusively, an even flatter
spectrum (say $\alpha_{\rm inj} \approx 2.2$) can be expected due to
increasing escape probability with momentum
\citep{2003A&A...399..409E}.\\
{\bf Supernova Remnants} (SNR) are known to be able to produce flat
($\alpha_{\rm inj} \approx 2.4$) CR populations and they are believed
to be the main CRp source of our galaxy \citep[][ and references
therein]{2002cra..book.....S}. 
The reason for the steeper ($\alpha_\p \approx 2.7$) galactic CRp spectrum
is thought to be the momentum dependent escape probability from our Galaxy.
Thus, the spectrum of CRp escaping
from galaxies should be flat again ($\alpha_{\rm inj}
\approx 2.4$). The spectrum injected into galaxy clusters could be
even flatter, if termination shock waves of the galactic winds are
able to re-accelerate them, as proposed by
\citet{1996SSRv...75..279V}.

The CRp population in galaxy clusters which is able to interact with
the thermal gas and thus to produce observable signatures
will be a mixture of contributions of the different sources,
modified by acceleration and energy loss processes. In order not to
rely too much on a specific physical picture, we discuss simplified
models, which should be able to capture many typical situations.

\subsection{Spatial distribution of cosmic ray protons in cooling flow
  clusters}  
\label{sec: CRp models}
In the following, we introduce three models for the spatial distribution of
  CRp within clusters of galaxies. The origin of the CRp population is not
  specified in the first two models, but the CRp may be accelerated by shock
  waves of cluster mergers, accretion shocks, or result from supernova driven
  galactic winds. In contrast to that we explore in the third model the
  diffusion process of CRp away from a central AGN. Since it is unclear how CRp
  are distributed spatially in detail, we investigate here three different
  scenarios which should serve as toy models. We pursue the philosophy of
  estimating physical parameters from observationally obtained electron density
  and temperature profiles by using simplified model assumptions for the CRp
  population. In doing so we try to minimize the dimensionality of parameter
  space as much as possible in order to track the main physical processes by
  means of analytically feasible methods and not to rely upon too many
  assumptions. Therefore the presented CRp profiles which are based on the
  assumption of spherical symmetry should not be interpreted as a precise
  estimate of the CRp population but rather as a plausible spherically averaged
  scenario.  

\subsubsection{Isobaric model of CRp}
\label{sec: isobaric}
In this model we assume that the average  kinetic CRp
energy density $\eps_{\crp} (\vec{r})$ is a constant fraction of the
thermal energy density $\eps_\mathrm{th} (\vec{r})$ of the ICM
\begin{equation}
\eps_\crp(\vec{r}) =  X_\crp \, \eps_\mathrm{th}(\vec{r})\,.
\end{equation}
This distribution might be maintained even in the case of a cooling flow
cluster by mixing and ongoing turbulent CRp diffusion processes exerted by
relativistic plasma bubbles rising in the gravitational potential of the
cluster due to buoyant forces 
\citep[][ and references therein]{2001ApJ...554..261C} which possibly leads to 
fractional pressure equilibrium with the thermal particle population.

\subsubsection{Adiabatic compression of CRp}
\label{sec: adiabatic}
Here we assume the CRp population to be originally isobaric to the thermal
population but to become adiabatically compressed during the
formation of the cooling flow while it did not relax afterwards. 
The phase space volume stays constant
during this transition and the momenta and volumes scale according to
\begin{eqnarray}
p_\p \to p'_\p &=& \left( \frac{n'_\e}{n_\e} \right)^{1/3}\!
                       p_\p = C^{1/3} p_\p\,, \\
V_\p \to V'_\p &=& \left( \frac{n'_\e}{n_\e} \right)^{-1} \!
                       V_\p = C^{-1} V_\p \,.
\end{eqnarray}
Here the compression factor $C=C(\vec{r}) = (n'_\e /n_\e)(\vec{r}) $ has
been introduced, which is larger than unity within cooling flows. 
Provided that the electrons have been in hydrostatic equilibrium during this
transition, this implies $C(\vec{r}) = T_\mathrm{cluster}/T_\e'(\vec{r})$, 
where $T_\mathrm{cluster}$ denotes the electron temperature in the outer core
region.
This transformation implicitly assumes that the ratio of the CRp number
densities before and after the adiabatic compression equals that of the
electron population.
If the differential number density distribution of the CRp population may be 
described by a power-law in momentum $p_\p$, then after adiabatic compression
of CRp the functional shape of their distribution remains unchanged, however
shifted according to 
\begin{eqnarray}
f'(\vec{r}',p') &=& \frac{\tilde{n}_\crp'(\vec{r}')\,c}{\mbox{GeV}}\,
\left( \frac{p'c}{\mbox{GeV}} \right)^{-\alpha_\p}\,, \\
\tilde{n}_\crp'(\vec{r}') &=& \tilde{n}_\crp[\vec{r}'(\vec{r})]\,C(\vec{r}')^{(\alpha_\p + 2)/3}.
\end{eqnarray}

The normalization $\tilde{n}_\crp(\vec{r})$ is chosen in such a way that the kinetic CRp
energy density makes up a constant fraction of the thermal energy density prior
to cooling flow formation and is described by a scaling parameter $X_\crp$
\begin{equation}
\eps_\crp(\vec{r}) = X_\crp\,\eps_\mathrm{th}(\vec{r}) \;\to\;
\eps'_\crp(\vec{r}') = X'_\crp(\vec{r}')\,\eps_\mathrm{th}(\vec{r}')\,.
\end{equation}
After adiabatic compression of CRp this scaling parameter has thus changed
to
\begin{equation}
\label{Xcrp adiabatic}
X'_\crp(\vec{r}) = C^{(\alpha_\p + 2)/3}(\vec{r})\, X_\crp\,.
\end{equation}
Since any hadronically induced emissivity scales with $X'_\crp$
we obtain the following relation,
\begin{equation}
j^\mathrm{~adiabatic}(\vec{r}) = C^{(\alpha_\p + 2)/3}(\vec{r})\, 
j^\mathrm{~isobaric}(\vec{r})\,.
\end{equation}

\subsubsection{Diffusion of CRp away from a central AGN}
\label{sec: AGNdiffusion}

Many galaxy clusters -- especially those with a cooling flow -- harbor a
central galaxy, which often exhibits nuclear activity.  The relativistic plasma
bubbles produced by the AGN may contain relativistic protons, which can partly
escape into the thermal ICM.  Most of the CRp that have been injected into the
cluster center are either diffusively transported into the surrounding ICM
\citep[as assumed by][]{1998APh.....9..227C, 1999ApJ...525..603B} or form
relativistic bubbles which rise in the gravitational potential of the cluster
due to buoyant forces \citep[][ and references therein]{2001ApJ...554..261C}.
An argument in favor of a significant central CRp injection into the ICM is the
much more efficient escape of CRp from the magnetic confinement of the radio
plasma bubble during the very early stages due to the bubbles higher
geometrical compactness and and expected stronger turbulence level
\citep{2003A&A...399..409E}.  In addition to this, any galactic wind from a
central galaxy will also inject CRp into the cluster center.  In order to
  treat these diffusion processes analytically one has to distinguish between
  clusters containing a cooling flow region or not. In the first case CRp
  diffusion will shape their emission profiles owing to the peaked cooling
  flow profiles while the emission strength in non-cooling flow clusters is
  mainly governed by the effective injection timescale.  

{\bf Cooling flow clusters:} The transport of CRp through the ICM is diffusive,
with a diffusion coefficient $\kappa(r, p)$ which in general may depend on
  momentum and position. For illustration we use
\begin{equation}
\label{eq:diff}
\kappa(r,p) = \kappa_0 (r)\, \left( \frac{p\,c}{{\mbox{GeV}}}
\right)^{\alpha_{\rm diff}}\,,
\end{equation}
with $\kappa_0 \sim 10^{27 \ldots 30}\, \mbox{cm}^2\,\mbox{s}^{-1}$ being
plausible values. By using this ansatz, we ignore likely deviations of the
diffusion coefficient from eq.~(\ref{eq:diff}) in the mildly relativistic
regime because these CRp are also not constrained by observations of their
hadronic interactions.  The coefficient $\alpha_{\rm diff}$ describes the
momentum dependence of the diffusion and is expected to be $\alpha_{\rm diff}
\approx \frac{1}{3}$ for active CRp diffusion in a Kolmogorov-like small-scale
magnetic turbulence spectrum and $\alpha_{\rm diff} \approx 0$ for passive
advective transport in a turbulent flow. In the latter case $\kappa(p) = v_{\rm
  turb}\,\lambda_{\rm turb}/3 \sim 10^{29}\, {\rm cm^2\,s^{-1}}$, where $v_{\rm
  turb}\sim 100\,{\rm km/s}$ and $\lambda_{\rm turb} \sim 10$ kpc are the
turbulent velocity and coherence length, respectively.

In a stationary situation, which is a valid approximation for timescales longer
than the typical CRp diffusion timescale in the case of a stationary or
short-term intermittent CRp source, the CRp distribution functions is given by
\begin{eqnarray}
\label{eq:diffsol}
f_\p (r, p_\p)  &=& -\frac{Q(p)}{4\, \pi} \int_0^r
\frac{\dd r'}{\kappa(r',p)\, r'^2} = 
\frac{Q_\p(p)}{4\,\pi\, \kappa(p)\,r}\,, \\
\mbox{where~~}
Q_\p(p) &=& \frac{Q_{\p,0} \,c}{\mbox{GeV}}\, \left(
\frac{p_\p \,c}{\mbox{GeV}} \right)^{-\alpha_{\rm inj}}
\end{eqnarray}
is the averaged CRp injection rate of the central source. We assume it to be a
power-law in momentum with spectral index $\alpha_{\rm inj}$, which in general
is not identical to the spectral index of the CRp population within radio
plasma since the escape fraction is expected to depend on momentum
\citep{2003A&A...399..409E}.  For the last step in eq.~(\ref{eq:diffsol}) we
  assume for simplicity the diffusion coefficient to be independent of
  position. Possible models of spatial distributions for the diffusion
  coefficient depend strongly on many unknown quantities such as the dominant
  diffusion mechanism (active diffusion versus passive advective transport),
  the velocity field, the turbulence scale, and the topology of the magnetic
  field, only to mention a few. Therefore we are unable to guess a realistic
  profile for $\kappa_0(r)$ without enlarging the accessible parameter space
  tremendously.  However since we expect the diffusion coefficient not to
  change dramatically over the cooling flow scale and since the distribution
  function $f_\p (r, p_\p)$ is sufficiently steep in radius
  (eq.~(\ref{eq:diffsol})) our results should be approximately correct.  The
total CRp luminosity of the source can be estimated from eq.~(\ref{eq:diffsol})
to be
\begin{equation}
L_\crp = \frac{m_\p\,c^2\, Q_{\p,0}}{2(\alpha_{\rm inj}-1)}\,
\left(\frac{m_\p\,c^2}{\mbox{GeV}} \right)^{1-\alpha_{\rm
inj}} \mathcal{B}\left(\frac{\alpha_{\rm inj}-2}{2},
\frac{3-\alpha_{\rm inj}}{2}\right)\!.
\end{equation}

Within our model, the CRp distribution function within the thermal ICM can be
written as
\begin{equation}
\label{AGNdistribution}
f_\p (r, p_\p) = \frac{\tilde{n}_{\crp,0}\,\,c}{\mbox{GeV}}\, 
\left(\frac{r}{h_{70}^{-1}\mbox{ kpc}}\right)^{-1}\,
\left(\frac{p_\p \,c}{\mbox{GeV}} \right)^{-\alpha_\p} \,,
\end{equation}
where $\alpha_\p = \alpha_{\rm inj} + \alpha_{\rm
diff}$.\footnote{This seems to be in contradiction to the identity of
injection and equilibrium spectral index for a system without escape
claimed in Sect. \ref{sec:consrtalpha}. Formally, we had to include
particle escape from the galaxy cluster in order to be able to have a
finite steady state solution of the diffusion problem, as given by
eq. (\ref{eq:diffsol}). In the realistic case of a finite age of the
system the stationary solution is only approximately valid in the
center of the galaxy cluster. However, only there exists a
sufficiently high target density to detect the CRp
population. Therefore, although we use a poor description of the CRp
profile on large-scales, the estimated $\gamma$-ray fluxes should be
sufficiently accurate.}

In order to obtain a realistic estimate for the diffusion volume to be
  considered, the relevant length scale needs to be taken into account.  We
  define the characteristic scale $R_\mathrm{diff}$ by calculating the second
  moment of the time-dependent distribution function of the first particles
  released by the source, yielding
  \begin{equation}
    \label{eq:diffscale}
    R_\mathrm{diff} = \sqrt{2\, n_\mathrm{dim}\, t_\mathrm{inj}\, \kappa(p)}
    \simeq 80\, h_{70}^{-1/2}\mbox{kpc}\,
    \left(\frac{p\,c}{\mbox{GeV}}\right)^{\alpha_{\rm diff}/2}\,,
  \end{equation}
  where $n_\mathrm{dim}=3$ denotes the number of spatial dimensions.  Here we
  assume a typical lifetime of $t_\mathrm{inj} = 3\, h_{70}^{-1}$ Gyr and
  $\kappa_0 \simeq 10^{29} \mbox{ cm}^2\mbox{ s}^{-1}$.  Beyond this scale
  $R_\mathrm{diff}$ there can be a CRp population resulting from diffusion away
  from the central AGN which is however exponentially suppressed. Because the
  $\gamma$-ray luminosity resulting from hadronic CRp interactions scales as
  \begin{equation}
    \label{eq:peak}
    \mathcal{L}_\gamma \propto  
    4\, \pi\, \int \dd r\, r^2 \tilde{n}_\crp(r)\, n_\e(r) \propto 
    \int \dd r\, r^{1-3\beta}\,,
  \end{equation}
  we always obtain centrally peaked $\gamma$-ray profiles, since the cooling
  radius is smaller than the diffusion scale, $r_{c_1}< R_\mathrm{diff}$, and
  $\beta > 1/3$ within cooling flow regions (compare Table~\ref{tab: density}).
  Thus, the $\gamma$-ray luminosity is only weakly dependent on
  $R_\mathrm{diff}$ as long as it reflects the correct order of magnitude.  

In this work we constrain $\tilde{n}_{\crp,0}$ with the aid of $\gamma$-ray
observations of galaxy clusters.  From these constraints limits on the
averaged CRp luminosity escaping from the radio plasma of the central
galaxy can be derived using
\begin{eqnarray}
\frac{L_\crp}{\kappa_0} &=& \frac{4\,\pi\,m_\p\,c^2\,
\tilde{n}_{\crp,0}\, h_{70}^{-1}\mbox{ kpc}}{2\,(\alpha_{\rm inj}-1)}\,
\left(\frac{m_\p\,c^2}{\mbox{GeV}} \right)^{1-\alpha_{\rm inj}}
\nonumber\\ 
\label{LCRp}
&&\times\,\mathcal{B}\left(\frac{\alpha_{\rm inj}-2}{2},
\frac{3-\alpha_{\rm inj}}{2}\right)\,,
\end{eqnarray}
where we again ignored any possible low-energy spectral cutoff, since
it can be included a posteriori with the help of
Fig.~\ref{fig:energyratio}. As a rough estimate we find numerically
\begin{eqnarray}
L_\crp &=& L(\alpha_{\rm inj})\, 10^{43}\, 
h_{70}^{-1/2}\mbox{ erg}\mbox{ s}^{-1} \nonumber\\
&&\times\,\left( \frac{\kappa_0 }{10^{29}\,{\rm cm^2\,s^{-1}}} \right)
\,\left( \frac{\tilde{n}_{\crp,0} }{10^{-6}\,h_{70}^{1/2}{\rm cm^{-3}}} \right),
\end{eqnarray}
with $L(\alpha_{\rm inj}) = 6.1$, 2.2, 1.6, and 1.7 for
$\alpha_{\rm inj} = 2.1,$ 2.3, 2.5, and 2.7, respectively. 
In Sect~\ref{sec:Np0} we analyze these constraints for our cluster sample in 
more detail.

{\bf Non-cooling flow clusters:} In transforming the above considerations on
  diffusion length scales to the case of non-cooling flow clusters we point out
  the following differences: In non-cooling flow clusters the core radius is
  normally larger than the diffusion scale, $r_{c}> R_\mathrm{diff}$, over
  which the electron density varies only slightly. Thus, a stationary solution
  to the diffusion equation is not applicable in the case of a flat target
  profile. It follows that the volume integrated $\gamma$-ray spectrum does not
  depend on the diffusion coefficient but only on the injection time
  $t_\mathrm{inj}$ of CRp into the ICM of the cluster core. We therefore adopt
  a modification to the diffusion model for non-cooling flow clusters. The
  averaged CRp luminosity of the central galaxy reads in this context
  \begin{eqnarray}
  L_\crp &=& \frac{\tilde{N}_\crp}{t_\mathrm{inj}}\,
             \frac{m_\p\,c^2\,}{2\,(\alpha_{\rm inj}-1)}\,
  \left(\frac{m_\p\,c^2}{\mbox{GeV}} \right)^{1-\alpha_{\rm inj}}
  \nonumber\\ 
  \label{LCRp_pot}
  &&\times\,\mathcal{B}\left(\frac{\alpha_{\rm inj}-2}{2},
  \frac{3-\alpha_{\rm inj}}{2}\right)\,.
  \end{eqnarray}
  Here $\tilde{N}_\crp$ denotes the integrated number of CRp being injected
  into the ICM of the cluster and $\alpha_{\rm inj} = \alpha_\p$, because there
  is no diffusion induced spectral steepening simply due to the fact that the
  even more energetic CRp which are still significantly contributing to the
  $\gamma$-ray flux in the EGRET energy band are not able to leave the central
  core region within a reasonable timescale.  

\subsection{Constraining the population of CRp by the integrated flux of
  $\gamma$-rays in different clusters}

\subsubsection{Cluster sample}

\begin{table*}[t]
\caption[t]{Parameters of electron density profiles $n_\e(r)$ of our cluster
 sample (central densities $n_i$ are subject to different formulae (\ref{ne})
 and (\ref{ne deproj})). The cluster are ordered according to their property of
 being a cooling flow cluster (upper part) or a non-cooling flow cluster (lower
 part).}  
\vspace{-0.1 cm}
\begin{center}
\begin{tabular}{lcl@{ $\times$ }lcrcll@{ }c@{ }lclcl}
\hline \hline
\vphantom{\Large A}%
 & & \multicolumn{2}{c}{$n_1$} & \multicolumn{3}{c}{$r_{\mathrm{c}_1}$} & &
     \multicolumn{3}{c}{$n_2$} & \multicolumn{1}{c}{$r_{\mathrm{c}_2}$} & & & \\ 
\vphantom{\LARGE p}%
 Cluster & $z$ & 
     \multicolumn{2}{c}{[$h_{70}^{1/2}\mbox{ cm}^{-3}$]} & 
           \multicolumn{3}{c}{[$h_{70}^{-1}\mbox{ kpc}$]} &
           \multicolumn{1}{c}{$\beta_1$} & 
     \multicolumn{3}{c}{[$h_{70}^{1/2}\mbox{ cm}^{-3}$]} &
           \multicolumn{1}{c}{[$h_{70}^{-1}\mbox{ kpc}$]} & 
           \multicolumn{1}{c}{$\beta_2$} & Equation & 
           \multicolumn{1}{c}{References} \\
\hline 
\vphantom{\Large A}%
\object{A85} \dotfill            & 0.0551 & $3.08$  & $10^{-2}$ && 45 && 0.662 
                                 & $3.87$ & $\times$ & $10^{-3}$ & 226  & 0.662  &
                                 (\ref{ne deproj})  & (a), (b)\\
\object{A426} (Perseus) \dotfill & 0.0179 & $4.6$   & $10^{-2}$ && 57 && 1.2 
                                 & $4.79$ & $\times$ & $10^{-3}$ & 200  & 0.58  & 
                                 (\ref{ne}) & (c), (d) \\
\object{A2199} \dotfill          & 0.0302 & $3.37$  & $10^{-2}$ && 29 && 0.663
                                 & $7.17$ & $\times$ & $10^{-3}$ & 116  & 0.663  &
                                 (\ref{ne deproj})  & (a), (b) \\
\object{A3526} (Centaurus)       & 0.0114 & $8.05$  & $10^{-2}$ && 8.6 && 0.569  
                                 & $3.65$ & $\times$ & $10^{-3}$ & ~~99 & 0.569  &
                                 (\ref{ne deproj})  & (a), (d) \\
\object{Ophiuchus} \dotfill      & 0.0280 & $1.71$  & $10^{-2}$ && 56 && 0.705 
                                 & $7.47$  & $\times$ & $10^{-3}$ & 190 & 0.705  &
                                 (\ref{ne deproj})  & (a), (e) \\
\object{Triangulum Australis}    & 0.0510 & $7.31$  & $10^{-3}$ && 151 && 0.816 
                                 & $2.63$ & $\times$ & $10^{-3}$ & 444 & 0.816  &
                                 (\ref{ne deproj})  & (a), (f) \\
\object{Virgo} \dotfill          & 0.0036 & $1.5$   & $10^{-1}$ && 1.6 && 0.42 
                                 & $1.3$  & $\times$ & $10^{-2}$ & ~~20   & 0.47 & 
                                 (\ref{ne deproj}) & (g), (h) \\                    
\hline 
\vphantom{\Large A}%
\object{A1656} (Coma) \dotfill   & 0.0231 & $3.4$ & $10^{-3}$  && 294 &&
                                 0.75  &&&&& & (\ref{ne}) & (i), (d) \\
\object{A2256} \dotfill          & 0.0581 & $3.57$ & $10^{-3}$ && 347 &&
                                 0.828 &&&&& & (\ref{ne}) & (a), (d) \\
\object{A2319} \dotfill          & 0.0557 & $7.35$ & $10^{-3}$ && 152 &&
                                 0.536 &&&&& & (\ref{ne}) & (a), (d) \\
\object{A3571} \dotfill          & 0.0391 & $9.37$ & $10^{-3}$ && 124 &&
                                 0.61  &&&&& &  
                                 (\ref{ne}) & (a), (d) \\
\hline 
\end{tabular}
\end{center}
\label{tab: density}

(a) \citet{1999ApJ...517..627M}, (b) \citet{2001AJ....122.2858O},
(c) \citet{2003ApJ...590..225C}, (d) \citet{1999ApJS..125...35S},
(e) \citet{1989MNRAS.238..881L}, (f) \citet{1981MNRAS.197..893M},
(g) \citet{2002A&A...386...77M}, (h) \citet{1998MNRAS.301..881E},
(i) \citet{1992A&A...259L..31B} 
\end{table*}

\begin{table*}[t]
\caption[t]{Parameters of temperature profiles $T_\e(r)$ of our cluster
 sample. The estimated $\gamma$-ray flux $\mathcal{F}_{\gamma,\,\mathrm{est}}\,
 (>100\mbox{ MeV})$ was calculated using the 
 $\mathcal{F}_{\gamma}$--$F_\mathrm{X}$ scaling relation
 (eq. (\ref{scaling})) with $\alpha_\p = 2.3$ and bolometric X-ray fluxes from
 \citet{1993ApJ...412..479D}. Note that $\mathcal{F}_{\gamma}$
 scales linearly with $X_\crp$ which was set to $X_\crp = 0.01\, X_{0.01}$ in this
 table. The range for $\mathcal{F}_{\gamma,\,\mathrm{est}}$ reflects the
 temperature spread in cooling flow clusters between the central temperature
 $T_0$ and the peripheral temperature $T_1$.}
\vspace{-0.1 cm}
\begin{center}
\begin{tabular}{llcccllcr@{~~~}ll}
\hline \hline
\vphantom{\Large Ap}%
& & \multicolumn{1}{c}{$k T_0$} & \multicolumn{1}{c}{$k T_1$} & 
     \multicolumn{2}{c}{$r_{\mathrm{temp}}$} &  &
     \multicolumn{3}{c}{$\mathcal{F}_{\gamma,\,\mathrm{est}}\,(>100\mbox{ MeV})$}
 & \\  
\vphantom{\LARGE $j_p$}%
 Cluster & Experiment & 
     \multicolumn{1}{c}{[keV]} & \multicolumn{1}{c}{[keV]} &
     \multicolumn{2}{c}{[$h_{70}^{-1}\mbox{ kpc}$]} & 
     \multicolumn{1}{c}{$\eta$} & 
     \multicolumn{3}{c}{$[X_{0.01}\,10^{-10} \mbox{ cm}^{-2} 
                         \mbox{ s}^{-1}]$} & 
     \multicolumn{1}{c}{References} \\
\hline 

\vphantom{\Large A}%
\object{A85} \dotfill            & BeppoSAX MECS & ~~5.5 & 9.0 && 312 & 2 &&
                                 1.8 & $\ldots$~~~2.3 & (a), (b) \\
\object{A426} (Perseus) \dotfill & XMM-Newton MOS & ~~3.0 & 7.0 && ~~94 & 3 &&
                                 19.1 & $\ldots$~29.2 & (c) \\
\object{A2199} \dotfill          & Chandra ACIS & ~~1.6 & 4.3 && ~~21.5 & 1.8 &&
                                 1.3 & $\ldots$~~~2.2 & (d), (e) \\
\object{A3526} (Centaurus)       & ASCA GIS & ~~2.2 & 4.0 && ~~22 & 3 &&
                                 2.2 & $\ldots$~~~2.9 & (f) \\
\object{Ophiuchus} \dotfill      & ASCA GIS & 12.8 && & & &&
                                 22.0 &  & (f) \\
\object{Triangulum Australis}    & ASCA GIS & 10.3 && & & &&
                                 4.8 & & (f) \\
\object{Virgo} \dotfill          & XMM-Newton PN/MOS & ~~1.0 & 3.0 && ~~13.5 & 1 &&
                                 3.2 & $\ldots$~~~5.6  & (g) \\
\hline 
\vphantom{\Large A}%
\object{A1656} (Coma) \dotfill   & XMM-Newton MOS & ~~8.3 && & & &&
                                 13.1 &  & (h) \\
\object{A2256} \dotfill          & Chandra ACIS & ~~6.7 && & & &&
                                 2.3 &  & (i) \\
\object{A2319} \dotfill          & ASCA GIS & ~~9.7 && & & &&
                                 5.4 &  & (f) \\
\object{A3571} \dotfill          & ASCA GIS & ~~7.2 && & & &&
                                 4.5 &  & (f) \\
\hline 
\end{tabular}
\end{center}
\label{tab: temperature}

(a) \citet{2000ApJ...538..543I}, (b) \citet{2001A&A...368..440L},
(c) \citet{2003ApJ...590..225C}, (d) \citet{2002MNRAS.335L...7V},
(e) \citet{2002MNRAS.336..299J}, (f) \citet{2000MNRAS.312..663W},
(g) \citet{2002A&A...386...77M}, (h) \citet{2001A&A...365L..67A}, 
(i) \citet{2002ApJ...565..867S}
\end{table*}

Applying the $\mathcal{F}_\gamma$--$F_\mathrm{X}$ scaling relation
(eq.~(\ref{scaling})) and taking bolometric X-ray fluxes from
\citet{1993ApJ...412..479D} while fixing $X_\crp = 0.01\,X_{0.01}$ and
$\alpha_\p = 2.3$ we estimated $\gamma$-ray fluxes
$\mathcal{F}_{\gamma,\,\mathrm{est}} ({>100}\mbox{ MeV})$ for the spectral
sensitivity of EGRET in order to choose our cluster sample (see Table~\ref{tab:
  temperature}). Inferred values for the estimated $\gamma$-ray flux
$\mathcal{F}_{\gamma,\,\mathrm{est}}$ by means of the
$\mathcal{F}_\gamma$--$F_\mathrm{X}$ scaling relation sensitively depend on the
bolometric X-ray luminosity of the particular cluster such that values for
$\mathcal{F}_{\gamma,\,\mathrm{est}}$ in Table~\ref{tab: temperature} represent
a rough estimate.  A detailed modeling using density and temperature profiles
will be described later on in Sect.~\ref{sec: simulated gamma} in order to
obtain upper limits on the CRp population.  By comparing $\gamma$-ray fluxes
$\mathcal{F}_{\gamma}$ obtained from these two different methods we recognized
an inconsistency for the Virgo and Centaurus cluster: This discrepancy is
explained by a too small aperture of the X-ray experiments analyzed by
\citet{1993ApJ...412..479D} giving rise to an underestimation of the X-ray flux
of these two nearest clusters in our sample ($z_\mathrm{Virgo} = 0.0036$ and
$z_\mathrm{Centaurus} = 0.0114$) and therefore an underestimate of
$\mathcal{F}_{\gamma,\,\mathrm{est}}$ for these two clusters.  Moreover, we
noticed a systematic discrepancy of the order of 50\% between the different
methods in cooling flow clusters which is due to an insufficient accounting for
the radial temperature variation in eq.~(\ref{scaling}).

Parameters of electron density profiles $n_\e(r)$ of our cluster sample are
given in Table~\ref{tab: density} where the clusters are ordered according to
their property of containing a cooling flow (upper part) or not (lower part). 
Note that the parameters are subject to different formulae (\ref{ne}) and
(\ref{ne deproj}), 
\begin{eqnarray}
\label{ne}
n_\e (r) &=& \sum_{i=1}^2 n_i \, \left( 1+\frac{r^2}{r^2_{\mathrm{c}_i}}
\right)^{-3\,\beta/2}\!\!, \\
\label{ne deproj}
n_\e (r) &=& \,\left[\frac{\tilde{\Lambda}[T_\e(0)]}{\tilde{\Lambda}[T_\e(r)]} 
\times\sum_{i=1}^2 n_i^2 \, \left( 1+\frac{r^2}{r^2_{\mathrm{c}_i}}
\right)^{-3 \,\beta} \right]^{1/2}\!.
\end{eqnarray}
The last equation (\ref{ne deproj}) follows from deprojection of X-ray surface
brightness profiles which are represented by double $\beta$ models. The
derivation of this deprojection is given in Appendix~\ref{sec: deprojection}.
For simplicity and consistency with the X-ray surface brightness profiles given
in \citet{1999ApJ...517..627M} we ignored the weak dependency on
$T_\e(r)$ in eq.~(\ref{ne deproj}).

In order to model the temperature profiles $T_\e (r)$ for our cooling flow
cluster sample we applied the universal temperature profile for relaxed
clusters proposed by \citet{2001MNRAS.328L..37A} to data taken from the
literature, 
\begin{equation}
\label{Te}
T_\e (r) = T_0 + (T_1 - T_0)\,
\left[ 1+\left( \frac{r}{r_\mathrm{temp}} \right)^{-\eta}\right]^{-1}.
\end{equation}
This equation matches the temperature profile well up to radii of $\sim
0.3\,r_\mathrm{vir}$, which is sufficient for our purposes since we are especially
interested in the core region of clusters. 
The parameters of the temperature profile for particular cluster are given in
Table~\ref{tab: temperature}.

\subsubsection{Simulated $\gamma$-ray flux normalized by EGRET limits: The case
  of Perseus cluster}
\label{sec: simulated gamma}

The volume integrated omnidirectional differential $\gamma$-ray source function
$Q_\gamma(E_\gamma)$ can be obtained by integrating eq.~(\ref{q gamma}).  We
integrated the volume out to a radius of $3\,h_{70}^{-1}$~Mpc which corresponds
to the characteristic distance where the simple $\beta$-model of electron
densities breaks down due to accretion shocks in clusters.  The integration
kernel $q_\gamma(\vec{r},E_\gamma)$ scales linearly with $\tilde{n}_\crp(\vec{r})$ (as
shown in eq.~(\ref{q gamma})) which is obtained by solving eqs.~(\ref{Xcrp
  scaling}) and (\ref{eps thermal}).  By comparing the integrated $\gamma$-ray
flux above 100~MeV, $\mathcal{F}_\gamma(>100 \mbox{ MeV})$, to EGRET upper
limits \citep[see][]{2003ApJ...588..155R}, we constrain the CRp scaling
parameter $X_\crp$. The inferred value for $X_\crp$ in the Perseus cluster
normalizes the differential $\gamma$-ray flux
\begin{equation}
\frac{\dd \mathcal{F}_\gamma}{\dd E_\gamma} \equiv 
\frac{Q_\gamma (E_\gamma)}{4 \pi D^2}
\end{equation}
in Fig.~\ref{fig:qgamma}. The $\pi^0$-meson decay induced distinct spectral
signature resulting in the peak at a $\gamma$-ray energy of $m_{\pi^0}\,c^2/2
\simeq 67.5$~MeV can be clearly seen. 

Figure~\ref{fig:qgamma} shows also upper limits on the differential
$\gamma$-ray flux owing to IC emission of hadronically originating CRe
represented by power-laws. 
The IC spectra are computed by means of eq.~(\ref{IC}) for different spectral
indices $\alpha_\p$ and zero magnetic field. 
Non-zero magnetic fields can be included since the IC spectra scale according
to $B^2_\mathrm{CMB}/(B(\vec{r})^2 + B^2_\mathrm{CMB})$ (see eq.~(\ref{Ne}))
which results in a lower normalization.

\subsubsection{Results on the scaling parameter $X_\crp$ using $\gamma$-ray
  observations in different clusters}

\begin{table*}[t]
\caption[t]{Upper limits on the CRp scaling parameter $X_\crp$ by comparing the
  integrated flux above 100 MeV to EGRET upper limits assuming a $\gamma$-ray
  spectral index in Dermer's model $\alpha_\gamma=\alpha_\p$. The spatial
  distribution of CRp is given by the isobaric and the adiabatic model of CRp,
  respectively (see Sects.~\ref{sec: isobaric} and \ref{sec: adiabatic}).}    
\vspace{-0.1 cm}
\begin{center}
\begin{tabular}{lc*{8}{c}}
\hline \hline
\vphantom{\Large A}%
& \multicolumn{1}{c}{$\mathcal{F}_{\gamma}\,(>100\mbox{ MeV})$}
& \multicolumn{4}{c}{$X_\crp^\mathrm{isobaric}$}
& \multicolumn{4}{c}{$X_\crp^\mathrm{adiabatic}$} \\
\vphantom{\LARGE p}%
Cluster &
  \multicolumn{1}{c}{$[10^{-8} \mbox{ cm}^{-2}\mbox{ s}^{-1}]$}
    & $\alpha_\p = 2.1$ & $\alpha_\p = 2.3$ & $\alpha_\p = 2.5$ & $\alpha_\p = 2.7$ 
    & $\alpha_\p = 2.1$ & $\alpha_\p = 2.3$ & $\alpha_\p = 2.5$ & $\alpha_\p = 2.7$ \\
\hline 

\vphantom{\Large A}%
A85 \dotfill            & $< 6.32$ & 3.53 & 1.97 & 2.09 & 3.11 & 2.58 & 1.41 & 1.48 & 2.16 \\
A426 (Perseus) \dotfill & $< 3.72$ & 0.14 & 0.08 & 0.08 & 0.13 & 0.12 & 0.06 & 0.07 & 0.10 \\
A2199 \dotfill          & $< 9.27$ & 6.14 & 3.42 & 3.64 & 5.42 & 5.74 & 3.18 & 3.38 & 5.00 \\
A3526 (Centaurus)       & $< 5.31$ & 1.54 & 0.86 & 0.91 & 1.36 & 1.45 & 0.80 & 0.85 & 1.26 \\
Ophiuchus \dotfill      & $< 5.00$ & 0.30 & 0.17 & 0.18 & 0.26 & & & & \\
Triangulum Australis    & $< 8.13$ & 1.93 & 1.07 & 1.14 & 1.70 & & & & \\
Virgo \dotfill          & $< 2.18$ & 0.18 & 0.10 & 0.11 & 0.16 & 0.16 & 0.09 & 0.09 & 0.14 \\
\hline                                                                     
\vphantom{\Large A}%
A1656 (Coma) \dotfill   & $< 3.81$ & 0.45 & 0.25 & 0.27 & 0.40 & & & & \\
A2256 \dotfill          & $< 4.28$ & 3.15 & 1.75 & 1.87 & 2.78 & & & & \\
A2319 \dotfill          & $< 3.79$ & 0.86 & 0.48 & 0.51 & 0.76 & & & & \\
A3571 \dotfill          & $< 6.34$ & 1.85 & 1.03 & 1.09 & 1.63 & & & & \\
                                 
\hline 
\end{tabular}
\end{center}
\label{tab: Xcrp gamma}
\end{table*}

\begin{table*}[t]
\caption[t]{{\bf Cooling flow clusters:} Upper limits on the CRp density
  parameter $\tilde{n}_{\crp,0}$ and average CRp luminosity $L_\crp$ of the
  central active galaxy by comparing the integrated flux above 100 MeV to EGRET
  upper limits assuming a $\gamma$-ray spectral index in Dermer's model
  $\alpha_\gamma=\alpha_\p$. The spatial distribution of CRp is is calculated
  according to the diffusion model of CRp away from a central AGN assuming
  $\alpha_\p = \alpha_\mathrm{inj} + \alpha_\mathrm{diff}$, where
  $\alpha_\mathrm{diff} = 1/3$. 
  {\bf Non-cooling flow clusters:} Upper limits on the CRp number parameter
  $\tilde{N}_\crp$ and average CRp luminosity $L_\crp$ without any diffusion
  induced spectral steepening, i.e.~$\alpha_\p = \alpha_\mathrm{inj}$. Note
  that $L_\crp$ scales in the case of cooling flow clusters with the diffusion
  coefficient $\kappa_0$ while it only depends on the CRp injection time
  $t_\mathrm{inj}$ for non-cooling flow clusters
  (see Sect.~\ref{sec: AGNdiffusion}).}
\vspace{-0.1 cm}
\begin{center}
\begin{tabular}{lc*{8}{c}}
\hline \hline
\vphantom{\LARGE A}%
& \multicolumn{4}{c}{$\tilde{n}_{\crp,0}~[h_{70}^{1/2}\mbox{ cm}^{-3}]$} 
& \multicolumn{4}{c}{$L_\crp~\left[h_{70}^{-1/2}\mbox{ erg}\mbox{ s}^{-1}
  \,\left( \frac{\kappa_0}{10^{29}\, \mathrm{cm}^2\, \mathrm{s}^{-1}} 
    \right)\right]$} \\
\vphantom{\LARGE p}%
CF Cluster
    & $\alpha_\p = 2.4$ & $\alpha_\p = 2.5$ & $\alpha_\p = 2.7$ & $\alpha_\p = 2.9$
    & $\alpha_\p = 2.4$ & $\alpha_\p = 2.5$ & $\alpha_\p = 2.7$ & $\alpha_\p = 2.9$ \\
\hline 
\vphantom{\Large A}%
A85 \dotfill            & $6.0 \times 10^{-5}$ & $7.2 \times 10^{-5}$ & $9.9 \times 10^{-5}$ & $1.3 \times 10^{-4}$
                        & $5.5 \times 10^{45}$ & $2.7 \times 10^{45}$ & $1.9 \times 10^{45}$ & $2.0 \times 10^{45}$ \\
A426 (Perseus) \dotfill & $2.4 \times 10^{-6}$ & $2.9 \times 10^{-6}$ & $3.9 \times 10^{-6}$ & $5.2 \times 10^{-6}$
                        & $2.2 \times 10^{44}$ & $1.1 \times 10^{44}$ & $7.4 \times 10^{43}$ & $8.1 \times 10^{43}$ \\
A2199 \dotfill          & $3.2 \times 10^{-5}$ & $3.8 \times 10^{-5}$ & $5.3 \times 10^{-5}$ & $7.0 \times 10^{-5}$
                        & $3.0 \times 10^{45}$ & $1.4 \times 10^{45}$ & $9.9 \times 10^{44}$ & $1.1 \times 10^{45}$ \\
A3526 (Centaurus)       & $4.3 \times 10^{-6}$ & $5.1 \times 10^{-6}$ & $7.1 \times 10^{-6}$ & $9.4 \times 10^{-6}$
                        & $3.9 \times 10^{44}$ & $1.9 \times 10^{44}$ & $1.3 \times 10^{44}$ & $1.5 \times 10^{44}$ \\
Ophiuchus \dotfill      & $1.5 \times 10^{-5}$ & $1.8 \times 10^{-5}$ & $2.5 \times 10^{-5}$ & $3.3 \times 10^{-5}$
                        & $1.4 \times 10^{45}$ & $6.6 \times 10^{44}$ & $4.6 \times 10^{44}$ & $5.1 \times 10^{44}$ \\
Triangulum Australis    & $1.4 \times 10^{-4}$ & $1.7 \times 10^{-4}$ & $2.3 \times 10^{-4}$ & $3.1 \times 10^{-4}$
                        & $1.3 \times 10^{46}$ & $6.2 \times 10^{45}$ & $4.3 \times 10^{45}$ & $4.8 \times 10^{45}$ \\
Virgo \dotfill          & $2.5 \times 10^{-7}$ & $3.1 \times 10^{-7}$ & $4.2 \times 10^{-7}$ & $5.6 \times 10^{-7}$
                        & $2.4 \times 10^{43}$ & $1.1 \times 10^{43}$ & $7.9 \times 10^{42}$ & $8.7 \times 10^{42}$ \\
\hline                                                                                 
\vphantom{\LARGE A}%
& \multicolumn{4}{c}{$\tilde{N}_{\crp}~[h_{70}^{1/2}]$} 
& \multicolumn{4}{c}{$L_\crp~\left[h_{70}^{3/2}\mbox{ erg}\mbox{ s}^{-1}
  \,\left( \frac{t_\mathrm{inj}}{3\, \mathrm{Gyr}}\right)^{-1}\right]$} \\
\vphantom{\LARGE p}%
NCF Cluster
    & $\alpha_\p = 2.1$ & $\alpha_\p = 2.3$ & $\alpha_\p = 2.5$ & $\alpha_\p = 2.7$
    & $\alpha_\p = 2.1$ & $\alpha_\p = 2.3$ & $\alpha_\p = 2.5$ & $\alpha_\p = 2.7$ \\
\hline 
\vphantom{\Large A}%
A1656 (Coma) \dotfill   & $1.6 \times 10^{64}$ & $2.5 \times 10^{64}$ & $3.7 \times 10^{64}$ & $5.1 \times 10^{64}$
                        & $2.7 \times 10^{45}$ & $1.5 \times 10^{45}$ & $1.6 \times 10^{45}$ & $2.4 \times 10^{45}$ \\
A2256 \dotfill          & $1.1 \times 10^{65}$ & $1.8 \times 10^{65}$ & $2.6 \times 10^{65}$ & $3.6 \times 10^{65}$
                        & $1.9 \times 10^{46}$ & $1.1 \times 10^{46}$ & $1.1 \times 10^{46}$ & $1.7 \times 10^{46}$ \\
A2319 \dotfill          & $4.5 \times 10^{64}$ & $7.0 \times 10^{64}$ & $1.0 \times 10^{65}$ & $1.4 \times 10^{65}$
                        & $7.5 \times 10^{45}$ & $4.2 \times 10^{45}$ & $4.4 \times 10^{45}$ & $6.6 \times 10^{45}$ \\
A3571 \dotfill          & $2.8 \times 10^{64}$ & $4.4 \times 10^{64}$ & $6.5 \times 10^{64}$ & $9.0 \times 10^{64}$
                        & $4.7 \times 10^{45}$ & $2.6 \times 10^{45}$ & $2.8 \times 10^{45}$ & $4.2 \times 10^{45}$ \\
                                 
\hline 
\end{tabular}
\end{center}
\label{tab: Np0}
\end{table*}

By employing the technique described in Sect.~\ref{sec: simulated gamma} we
constrained the CRp scaling parameter $X_\crp$ using EGRET upper limits of the
$\gamma$-ray flux by \citet{2003ApJ...588..155R}.  As described in that
section, we infer the $\gamma$-ray flux of this clusters originating from
within a sphere of radius $3\,h_{70}^{-1}$~Mpc.  Owing to the vicinity of the
Virgo cluster this maximum radius subtends an angle on the sky which is larger
than the width of the point spread function of the EGRET instrument
($\theta_\mathrm{max} = 5.8\degr\,[E_\gamma/(100\mbox{ MeV})]^{-0.534}$,
\citet{2003ApJ...588..155R}). Thus in the case of Virgo we use this smaller
integration volume.  Table~\ref{tab: Xcrp gamma} shows constraints for $X_\crp$
using the isobaric and the adiabatic model of CRp described in Sect.~\ref{sec:
  CRp models}. Because in the adiabatic model the CRp scaling parameter
$X_\crp$ is a function of radius, the value $X_\crp^\mathrm{adiabatic}$ given
in Table~\ref{tab: Xcrp gamma} refers to the unprimed quantity in
eq.~(\ref{Xcrp adiabatic}) which reflects the outer core region of the cluster.
For clusters like Perseus (A~426), Virgo, Ophiuchus, and Coma (A~1656) we can
obtain quite tight constraints on the population of CRp.

\subsubsection{Results on $L_\crp$ in the AGN-diffusion model}
\label{sec:Np0}

The procedure of inferring constraints on CRp diffusing away from a central
source is mostly sensitive to the CRp population of the central cooling flow
region rather than the shock region in the outer parts of the cluster.  In
order to constrain the CRp density parameter $\tilde{n}_{\crp,0}$ and averaged
CRp luminosity $L_\crp$ of the central active galaxy in our AGN-diffusion model
of {\em cooling flow clusters} we have to calculate the volume integrated
omnidirectional differential $\gamma$-ray source function $Q_\gamma(E_\gamma)$
(see eq.~(\ref{Q})).  The integration kernel $q_\gamma(E_\gamma)$ is
proportional to $\tilde{n}_\crp(\vec{r})$ (eq.~(\ref{q gamma})) which is
obtained by solving eqs.~(\ref{fp}) and (\ref{AGNdistribution}).  By comparing
the integrated $\gamma$-ray flux above 100~MeV to EGRET upper limits
\citep[see][]{2003ApJ...588..155R}, we constrain the CRp density parameter
$\tilde{n}_{\crp,0}$.  In the case of {\em non-cooling flow clusters} we
constrain the averaged CRp luminosity $L_\crp$ with the aid of the integrated
CRp number parameter $\tilde{N}_\crp$, yielding an indirect measure of a
combination of the CRp escape fraction from the radio plasma of the central
galaxy and the averaged CRp luminosity of this source.

Upper limits on the CRp density parameter $\tilde{n}_{\crp,0}$, number
  parameter of CRp $\tilde{N}_\crp$, and averaged CRp luminosity $L_\crp$
of the central active galaxy (by means of eq.~(\ref{LCRp})) are presented in
Table~\ref{tab: Np0}.  This shows that within this conceptually simple model we
are able to put constraints on the averaged CRp luminosity $L_\crp$.
The limits which are strongest in the case of \object{M87} in the Virgo
  cluster represent conservative bounds since we choose the active CRp
  diffusion scenario resulting in spectral steepening of the CRp population. We
  obtain even tighter limits when assuming a passive advective transport of the
  CRp in a turbulent flow in which case we infer
  \begin{equation}
  L_\crp = L(\alpha_\p)\,10^{42}\, h_{70}^{-1/2}\mbox{ erg}\mbox{ s}^{-1}\,
  \left( \frac{\kappa_0 }{10^{29}\,{\rm cm^2\,s^{-1}}} \right)
  \end{equation}
  with $L(\alpha_\p)= 4.5,$ 4.8, 7.2, and 20.9 for $\alpha_\p = 2.4,$ 2.5, 2.7,
  and 2.9, respectively.  These values are slightly smaller than
  instantaneous jet power estimates of M87 being of the order of
  $L_\mathrm{jet} \simeq 10^{43}\mbox{ erg}\mbox{ s}^{-1}$
  \citep{1996ApJ...467..597B,2002ApJ...579..560Y}.  In general, this
  demonstrates the ability of future high resolution $\gamma$-ray observations
  to constrain the energy fraction of CRp escaping from the radio plasma.  

\subsection{Radio emissivity of secondary electrons: The case of
  the radio mini-halo in Perseus} 

In contrast to $\gamma$-rays induced by hadronic CRp interactions whose
spectral shape and normalization is only governed by the spectral index
$\alpha_\p$ as free parameter, the resulting radio emission from secondary
electrons also depends on the morphology and strength of the magnetic field
$B(\vec{r})$.
Because only a subsample of cooling flow clusters contain radio mini-halos
which are not outshined by the central AGN we decided to concentrate on the
Perseus cluster. It has the fortunate property that the radio emission due to
the central galaxy \object{NGC 1275} is spatially resolved and can be separated
from the diffuse emission due to the radio-mini halo.

\subsubsection{Intracluster magnetic fields}

Magnetic fields in galaxy clusters seem to be on the level of $\sim
\mu$G. Indirect estimates of magnetic field strength assuming
equipartition of energy density of the fields and that of a
radio synchrotron emitting relativistic electron population
give low field strengths of $\sim 0.1\,\mu$G. Also lower limits on the
field strength of a comparable level can be derived using the
measurements or upper limits on IC scattered CMB photons in the hard
X-ray band \citep{1994ApJ...429..554R, 1999ApJ...513L..21F,
1999AA...344..409E}. Conversely, Faraday rotation measurements indicate
magnetic fields strengths of several $\mu$G in typical galaxy clusters
and a few $10\,\mu$G in cooling flow regions of clusters \citep[][ for
a review]{2002ARA&A..40..319C}. Faraday rotation based measurements of
the field strength depend on estimating the magnetic autocorrelation  
length from fluctuations in the Faraday rotation maps. Although the   
formerly used methods to estimate this length-scale seem to be
questionable \citep{2003A&A...401..835E} a refined analysis gives
comparable results for the magnetic field strengths (Vogt \&
En{\ss}lin 2003, in preparation).

\subsubsection{Comparison of the morphology of radio emissivity from secondary
  electrons} 

\begin{figure}[t]
\resizebox{\hsize}{!}{
\includegraphics{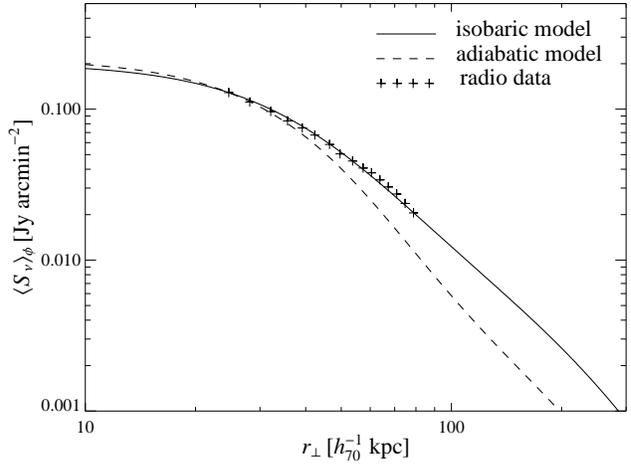}}
\caption{The radial distribution of radio brightness as a function of impact
  parameter $r_\bot$. Shown are the CRp adiabatic and isobaric model
  for model parameters $B_0 = 10 ~\mu \mbox{G}$, $\alpha_B = 0.5$, and
  $\alpha_\p = 2.3$ (details are described in the text) as well as the
  azimuthally averaged radio brightness profile of the the Perseus radio-mini
  halo \citep[data was taken from][]{1990MNRAS.246..477P}.
  The normalization of the radio brightness depends on the assumed scaling
  between CRp and thermal energy density. We fix this scaling parameter $X_\crp$
  by comparing the simulated radio brightness to the measured data at
  24.65~$h_{70}^{-1}$~kpc.}
\label{fig:Snu}
\end{figure}

The radio data was taken from \citet{1990MNRAS.246..477P} where we neglected
the innermost data points because of enhanced contribution to radio brightness
of the radio jet of \object{NGC 1275} and the outermost data points due to the
limited sensitivity on the larger scales of the specific VLA configuration
likely leading to an artificial decline in the radio surface brightness.  The
values for the azimuthally averaged radio surface brightness were converted
assuming a two-dimensional Gaussian beam which leads to a beam area
$A_\mathrm{beam} = \pi\,(4\,\ln2)^{-1}\,\mbox{FWHM}_x\,\mbox{FWHM}_y$.
Figure~\ref{fig:Snu} shows the radial distribution of radio brightness $S_\nu
(r_\bot)$ as a function of impact parameter $r_\bot$ obtained by means of
eq.~(\ref{Snu}) in comparison to the radio data.  The CRp adiabatic and
isobaric model being described in Sect.~\ref{sec: CRp models} are both shown
using model parameters of $\alpha_\p = 2.3$, $B_0 = 10 ~\mu \mbox{G}$, and
$\alpha_B = 0.5$, where the latter two parameters refer to eq.~(\ref{magnetic
  field}).  The normalization of the radio brightness depends on the assumed
scaling between CRp and thermal energy density. We fix this scaling parameter
$X_\crp$ by comparing the simulated radio brightness to the measured data at
24.65~$h_{70}^{-1}$~kpc.  There is an excellent morphological concordance of
the isobaric model of CRp and the radio data for the radio-mini halo of the
Perseus cluster.  Since the required values of $X_\crp$ are plausible ($\sim$
0.01--0.1, see Sect.~\ref{sec:XCRp_radio}), the hadronic secondary CRe model
is a very attractive explanation for the observed radio mini-halos in cooling
flow clusters.

\subsubsection{Results on the scaling parameter $X_\crp$ using radio
  observations in different models}
\label{sec:XCRp_radio}

\begin{table}[t]
\caption[t]{Upper limits on the CRp scaling parameter $X_\crp$ inferred from
  radio brightness profiles of the radio mini-halo of Perseus cluster for
  different values of $B_0$, $\alpha_B$, and $\alpha_\p$.}   
\vspace{-0.1 cm}
\begin{center}
\begin{tabular}{cccccc}
\hline \hline
\vphantom{\LARGE $A_p$}%
Model & $\alpha_\p$ & $B_0~[\mu \mbox{G}]$ & $\alpha_B$
      & $X_\crp^\mathrm{isobaric}$ & $X_\crp^\mathrm{adiabatic}$ \\
\hline 
\vphantom{\Large A}%
1 & 2.3 &  10 & 0.5 & 0.016 & 0.006 \\
2 & 2.1 &  10 & 0.5 & 0.014 & 0.005 \\
3 & 2.5 &  10 & 0.5 & 0.033 & 0.011 \\
4 & 2.7 &  10 & 0.5 & 0.096 & 0.031 \\
5 & 2.3 & ~~5 & 0.5 & 0.027 & 0.009 \\
6 & 2.3 &  20 & 0.5 & 0.012 & 0.004 \\
7 & 2.3 &  10 & 0.7 & 0.017 & 0.006 \\
8 & 2.3 &  10 & 0.9 & 0.019 & 0.006 \\
                    
\hline 
\end{tabular}
\end{center}
\label{tab: Xcrp radio}
\end{table}

By comparing the simulated radio brightness to the measured radio data at
$24.65 \,h_{70}^{-1}\mbox{ kpc}$ which is the innermost azimuthally averaged
data point not being outshined by the radio galaxy cocoon of \object{NGC 1275}
we determine the CRp scaling parameter $X_\crp$. Taking this point of reference 
yields more conservative upper limits for $X_\crp$
instead of normalizing by the integrated radio surface brightness especially in
the case of poorer morphological matches. The inferred values for $X_\crp$ in
Table~\ref{tab: Xcrp radio} are shown for different combinations of $B_0$,
$\alpha_B$, and $\alpha_\p$. 

Deduced values of this scaling parameter $X_\crp$ which are obtained by
considering only pion decay induced secondary electrons resulting from hadronic
CRp interactions in the ICM reflect upper limits because there are also other
mechanisms in galaxy clusters leading to relativistic populations of electrons
(see Sect.~\ref{sec:intro}). 
By analyzing the variations of our model parameters in Table~\ref{tab: Xcrp
  radio} we conclude a weak dependence of $X_\crp$ on $\alpha_B$ while the
magnetic field strength at the cluster center $B_0$ and the CRp spectral index
$\alpha_\p$ show a stronger influence on $X_\crp$.
The spectral parameter of the magnetic field $\alpha_B$ impacts mostly on the
radial extensions of the radio brightness profiles while the CRp scaling
parameter reflects a degeneracy with respect to $B_0$ and $\alpha_\p$.

\begin{figure}[t]
\resizebox{\hsize}{!}{
\includegraphics{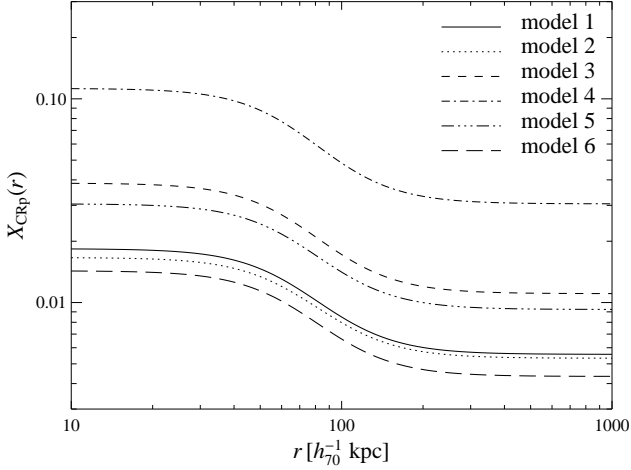}}
\caption{The deprojected scaling parameter $X_\crp (r)$ between CRp and thermal
  energy density in the adiabatic model applied to the mini-radio halo of
    Perseus and presented for models which are defined in Table~\ref{tab: Xcrp
    radio}.}
\label{fig:Xcrp} 
\end{figure}

Figure~\ref{fig:Xcrp} shows the scaling parameter $X_\crp (r)$ as a function of
radius $r$ between CRp and thermal energy density in the adiabatic model
according to eq.~(\ref{Xcrp adiabatic}) for models defined in Table~\ref{tab:
  Xcrp radio}.  The enhancement of CRp relative to the thermal energy density
owing to adiabatic compression of the CRp population during the formation of
the cooling flow can be clearly seen.

\subsection{Constraints derived from the radio halo of Coma}

\subsubsection{Parameter study of the hadronic scenario}
\label{sec:parameter}

\begin{figure}[t]
  \resizebox{\hsize}{!}{ \includegraphics{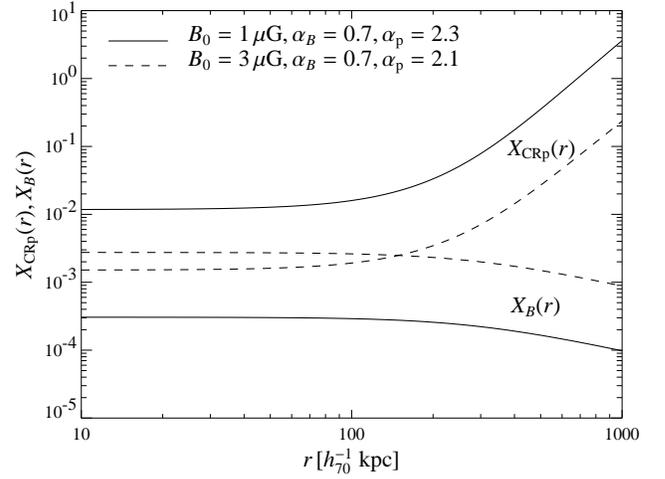}}
\caption{The deprojected CRp scaling parameter $X_\crp (r)$ required to
  account for the observed radio halo in Coma within the framework of the
  hadronic scenario. The rising curves with increasing radius represent
  $X_\crp (r)$ while the declining curves show $X_B(r)$ for the particular
  choice of a magnetic field being frozen into the flow and isotropized,
  i.e.~$\alpha_B = 0.7$ \citep{1993MNRAS.263...31T}.}
\label{fig:ComaXcrp} 
\end{figure}

We also applied this formalism of synchrotron radiation emitted by secondary
electrons as presented in Sect.~\ref{sec: radio} to the radial distribution of
radio brightness in the radio halo of the Coma cluster using radio data at
1.4~GHz by \citet{1997A&A...321...55D}.  Assuming the CRp population to be
distributed according to the isobaric model, the spatial radio brightness
profile obtained by this secondary electron model declines too fast with
increasing impact parameter $r_\bot$ in order to account for the observed
extended radio halo of Coma. To check whether this shortfall of the theoretical
model represents a serious problem for the hadronic model of radio synchrotron
emission we are asking in turn for the necessary radial variation of the CRp
scaling parameter $X_\crp(r)$ that is able to explain the observed radio halo.
Deprojecting the azimuthally averaged observed radio surface brightness profile
which is described by a $\beta$-model yields (as laid down in
Appendix~\ref{sec: deprojection})
\begin{equation}
  \label{eq:Coma:radio}
  j_\nu (r) = \frac{S_0}{2\pi\, r_\mathrm{c}}\, 
  \frac{6\beta - 1}{\left(1 + r^2/r_\mathrm{c}^2\right)^{3 \beta}}\,
  \mathcal{B}\left(\frac{1}{2}, 3\beta\right)\,,
\end{equation}
where $S_0 = 1.1 \mbox{ mJy arcmin}^{-2}$, $r_\mathrm{c} = 450\, h_{70}^{-1}
\mbox{ kpc}$, and $\beta = 0.78$. By comparing the observed to the
theoretically expected radio emissivity at each radius we infer the ratio of
CRp-to-thermal energy density $X_\crp(r)$. Figure~\ref{fig:ComaXcrp} shows a
comparison of $X_\crp(r)$ and the ratio of magnetic-to-thermal energy density
$X_B(r) = \eps_B(r)/\eps_\mathrm{th}(r)$ for particular model parameters
$\alpha_\p$, $\alpha_B$, and $B_0$. Whereas $\alpha_\p$ and $B_0$ impact mostly
on the normalization of both scaling parameters $X_\crp(r)$ and $X_B(r)$, the
choice of $\alpha_B$ governs the relative curvature of these functions:
$X_B(r)\propto n_\e(r)^{2\alpha_B - 1}$ is curved in a convex fashion for
$\alpha_B>0.5$ and exhibits concave curvature for $\alpha_B<0.5$ assuming the
cluster to be isothermal which is a valid approximation for Coma.  While there
are combinations of parameters for which $X_\crp(r)$ becomes larger than unity
and thus question the hadronic scenario \citep{2002BrunettiTaiwan}, only small
variations in parameter space yield plausible values for $X_\crp(r)$ (compare
Fig.~\ref{fig:ComaXcrp}).

\begin{figure}[t]
  \resizebox{\hsize}{!}{\includegraphics{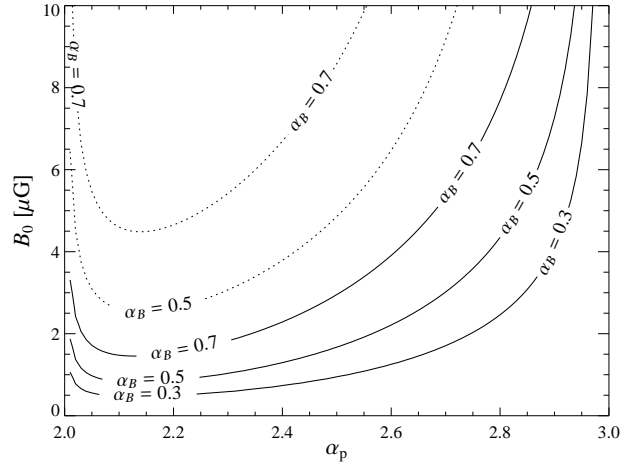}}
\caption{Parameter study on the ability of hadronically originating CRe to
  generate the radio halo of Coma.  Shown are contour lines of $X_\crp(r \le
  1\,h_{70}^{-1}\mbox{ Mpc}) = 1$ (solid) and $X_\crp(r \le
  1\,h_{70}^{-1}\mbox{ Mpc}) = 0.1$ (dotted) in parameter space spanned by
  $\alpha_\p$ and $B_0$ for three choices of magnetic field morphology
  characterized by $\alpha_B$. The contour line of $X_\crp = 0.1$ for $\alpha_B
  = 0.3$ has been omitted since it almost coincides with the contour of $X_\crp
  = 1$ for $\alpha_B = 0.7$.  The lower right corner represents the region in
  parameter space, where the hadronic scenario faces challenges for explaining
  the observed radio halo of Coma.}
\label{fig:parameter_space} 
\end{figure}

In order to quantify these considerations we perform a parameter study to
exclude regions of parameter space spanned by $\alpha_\p$, $\alpha_B$, and
$B_0$ where the hadronic scenario is challenged to account for the radio halo
in Coma.  Figure~\ref{fig:parameter_space} shows the resulting contour lines of
$X_\crp(r \le 1\,h_{70}^{-1}\mbox{ Mpc}) = 1$ and $X_\crp(r \le
1\,h_{70}^{-1}\mbox{ Mpc}) = 0.1$ in this parameter space. The gradient of
$X_\crp(r \le 1\,h_{70}^{-1}\mbox{ Mpc})$ points towards the lower right corner
in Fig.~\ref{fig:parameter_space} and thus leaves the upper left region of
parameter space where the hadronic scenario is able to account for the observed
radio halo depending on the specific choice of $\alpha_B$.  Since $X_B(r \le
1\,h_{70}^{-1}\mbox{ Mpc}) < 0.1$ for the entire region of parameter space
investigated here there are no further constraints imposed on the hadronic
scenario.

Choosing the energy density of the magnetic field to decline like the thermal
energy density, i.e.~$\alpha_B = 0.5$, requires $X_\crp(r)$ to increase by a
factor of less than two orders of magnitude from the center to the outer parts
of the cluster in order to reproduce the observed radio halo of Coma. This
factor, however, is reduced for smaller values of $\alpha_B$. It is further
reduced due to the non-spherical morphology of Coma, as explained in the
following.  The X-ray emissivity and the radio emissivity resulting from
hadronic CRp interactions differ in their scaling with the electron density
according to
\begin{eqnarray}
  \label{eq:anisotropy}
  \Lambda_\mathrm{X}(r)&\propto& n_\mathrm{e}(r)^2
  \quad\mbox{and}\quad \\
  j_\nu(r) &\propto& X_\crp(r)\, n_\mathrm{e}(r)^{2+\alpha_B(1 + \alpha_\p/2)}
  \sim X_\crp(r)\, n_\mathrm{e}(r)^{3\ldots 4}
\end{eqnarray}
within the framework set by our model and depending on the particular choice of
$\alpha_B$ and $\alpha_\p$. Thus, any anisotropy like the Coma X-ray and radio
bridge yields biased azimuthal averages when comparing observational to
theoretical radio surface brightness profiles where the latter uses density
profiles obtained by deprojecting X-ray profiles.  Remarkably, this discrepancy
is largest for large values of $\alpha_B$ and $\alpha_\p$ for which we infer
the tightest limits on the hadronic scenario
(cf.~Fig.~\ref{fig:parameter_space}) and thus softens these limits.  This
results in biased profiles of $X_\crp(r)$ which increase too strongly towards
larger radii \citep[cf.~also][]{2000A&A...362..151D}. Pursuing an approach of
averaging only along the line of sight could attenuate the bias
\citep{2001A&A...369..441G}.

An increase of $X_\crp(r)$ towards the cluster's periphery is indeed observed
in cosmological structure formation simulations due to adiabatic compression
inside the cluster which increases the thermal pressure at a higher rate than
the CRp pressure \citep{2001ApJ...562..233M, 2001ApJ...559...59M}. Bearing in
mind that the CRp-to-thermal pressure ratio of \citet{2001ApJ...562..233M,
  2001ApJ...559...59M} is obtained from volume averages and the energy density
stored in magnetic fields declines shallower in comparison to the thermal
energy density we conclude that our results arising the parameter study may be
well in agreement with these simulations.

\subsubsection{The spectrum of the Coma radio halo}

\begin{figure}[t]
  \resizebox{\hsize}{!}{ \includegraphics{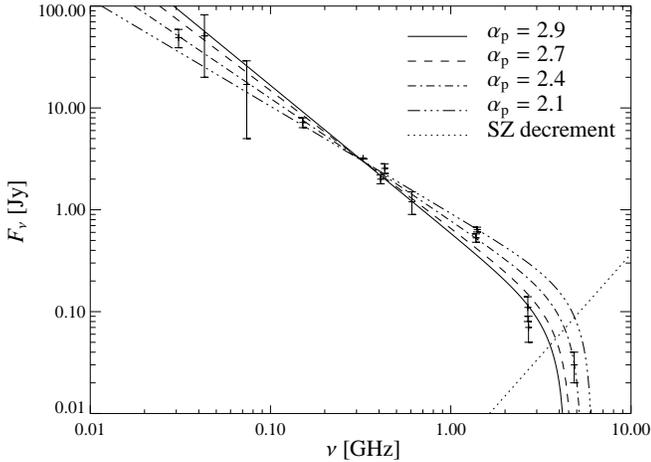}}
\caption{Observed radio halo fluxes of the Coma cluster as compiled by
    \citet{2003A&A...397...53T}. Shown are synchrotron power-law spectra for
    different spectral indices $\alpha_\nu = \alpha_\p/2$. The spectra are
    modified at higher frequencies by means of the SZ effect while the SZ
    decrement (a negative flux at the Rayleigh-Jeans part of the Planck
    spectrum) is also shown (dotted). The SZ flux is derived from the cluster
    volume up to the assumed position of the accretion shock.  }
\label{fig:Coma_spectrum} 
\end{figure}

One might object that the CRp spectral index should be determined better owing
to radio observations than the range of $\alpha_\p = (2,3)$ being considered in
the previous parameter study (Sect.~\ref{sec:parameter}).  The following line
of argumentation shows, that this, on the contrary, is not the case.  First,
there is an ambiguity of relating the CRp spectral index $\alpha_\p$ to the
induced synchrotron spectral index $\alpha_\nu$ which is either $\alpha_\nu =
\alpha_\p /2$ (Dermer's model) or $\alpha_\nu = (2\, \alpha_\p - 1)/3$
(fireball model). When comparing multifrequency observations of diffuse radio
emission of the ICM which extends to several GHz the Sunyaev-Zel'dovich (SZ)
distortion of the spectrum has to be taken care of. At these frequencies of the
Rayleigh-Jeans part of the Planck spectrum the SZ effect amounts to a decrement
which introduces a cutoff in the radio spectrum as can be seen in
Fig.~\ref{fig:Coma_spectrum}.  Following \citet{2002A&A...396L..17E} the SZ
luminosity reads for Coma in the Rayleigh-Jeans part
\begin{equation}
  \label{eq:SZ}
  F_\mathrm{SZ}^\mathrm{Coma} = -4.1 \times 10^{-3} \nu_\mathrm{GHz}^2\, 
  h_{70}^{-1/2}\mbox{ Jy}\,,
\end{equation}
where $\nu_\mathrm{GHz} = \nu/\mbox{GHz}$.\footnote{Here we corrected for a
  missing factor of 2 in equation (4) in \citet{2002A&A...396L..17E} and
  changed the slope of the $\beta$-profile to $\beta = 0.75$
  \citep{1992A&A...259L..31B}.}  However, the SZ amplitude is uncertain within
a factor of 2 which stems mostly from density profiles being inferred from
X-ray observations and extrapolated to $R_\mathrm{shock}$.  Furthermore, the
multifrequency dataset as compiled by \citet{2003A&A...397...53T} is
inhomogeneous because the solid angle over which the observed radio fluxes have
been integrated may vary among these observations. Finally, the quoted
uncertainties may underestimate the systematic uncertainties which e.g.~result
from incomplete accounting for point source subtraction.

\section{Detectability of $\gamma$-rays by satellite missions and \v{C}erenkov
  telescopes}

\begin{table*}[t]
\caption[t]{Expected limits on the CRp scaling parameter $X_\crp$ by comparing
  the integrated pion decay induced $\gamma$-ray flux above 100 GeV to
  sensitivity limits of \v{C}erenkov telescopes of 
  $\mathcal{F}_{\gamma,\,\mathrm{exp}}(E>E_\mathrm{thr}) = 
  10^{-12}\,\gamma \mbox{ cm}^{-2} \mbox{ s}^{-1}
  \, (E_\mathrm{thr}/100 \mbox{ GeV})^{1-\alpha_\gamma}$ 
  assuming a $\gamma$-ray spectral index in Dermer's model
  $\alpha_\gamma=\alpha_\p$.  
  Note that limits on $X_\crp$ roughly $\lesssim 0.01$ for $\alpha_\p = 2.3$ in the
  isobaric model provide good chances to detect $\gamma$-rays in these particular
  clusters with new generation \v{C}erenkov telescopes.}     
\vspace{-0.1 cm}
\begin{center}
\begin{tabular}{lccclcccl}
\hline \hline
\vphantom{\Large A}%
& \multicolumn{4}{c}{$X_\crp^\mathrm{isobaric}$}
& \multicolumn{4}{c}{$X_\crp^\mathrm{adiabatic}$} \\
\vphantom{\LARGE p}%
Cluster 
    & $\alpha_\p = 2.1$ & $\alpha_\p = 2.3$ & $\alpha_\p = 2.5$ & $\alpha_\p = 2.7$ 
    & $\alpha_\p = 2.1$ & $\alpha_\p = 2.3$ & $\alpha_\p = 2.5$ & $\alpha_\p = 2.7$ \\
\hline 

\vphantom{\Large A}%
A85 \dotfill            & $3.7 \times 10^{-2}$ & $6.5 \times 10^{-2}$ & $2.2 \times 10^{-1}$ & 1.0 
                        & $2.7 \times 10^{-2}$ & $4.7 \times 10^{-2}$ & $1.5 \times 10^{-1}$ & $7.1 \times 10^{-1}$ \\
A426 (Perseus) \dotfill & $2.5 \times 10^{-3}$ & $4.5 \times 10^{-3}$ & $1.5 \times 10^{-2}$ & $7.0 \times 10^{-2}$ 
                        & $2.1 \times 10^{-3}$ & $3.6 \times 10^{-3}$ & $1.2 \times 10^{-2}$ & $5.5 \times 10^{-2}$ \\
A2199 \dotfill          & $4.4 \times 10^{-2}$ & $7.7 \times 10^{-2}$ & $2.6 \times 10^{-1}$ & 1.2           
                        & $4.1 \times 10^{-2}$ & $7.2 \times 10^{-2}$ & $2.4 \times 10^{-1}$ & 1.1  \\
A3526 (Centaurus)       & $1.9 \times 10^{-2}$ & $3.4 \times 10^{-2}$ & $1.1 \times 10^{-1}$ & $5.3 \times 10^{-1}$ 
                        & $1.8 \times 10^{-2}$ & $3.2 \times 10^{-2}$ & $1.1 \times 10^{-1}$ & $5.0 \times 10^{-1}$ \\
Ophiuchus \dotfill      & $4.0 \times 10^{-3}$ & $7.0 \times 10^{-3}$ & $2.3 \times 10^{-2}$ & $1.1 \times 10^{-1}$ &&&& \\
Triangulum Australis    & $1.6 \times 10^{-2}$ & $2.8 \times 10^{-2}$ & $9.3 \times 10^{-2}$ & $4.4 \times 10^{-1}$ &&&& \\
Virgo \dotfill          & $4.1 \times 10^{-3}$ & $7.3 \times 10^{-3}$ & $2.4 \times 10^{-2}$ & $1.1 \times 10^{-1}$ 
                        & $3.8 \times 10^{-3}$ & $6.7 \times 10^{-3}$ & $2.2 \times 10^{-2}$ & $1.0 \times 10^{-1}$ \\
\hline                                                                                                  
\vphantom{\Large A}%
A1656 (Coma) \dotfill   & $7.9 \times 10^{-3}$ & $1.4 \times 10^{-2}$ & $4.6 \times 10^{-2}$ & $2.2 \times 10^{-1}$ &&&& \\
A2256 \dotfill          & $4.9 \times 10^{-2}$ & $8.6 \times 10^{-2}$ & $2.9 \times 10^{-1}$ & 1.4           &&&& \\
A2319 \dotfill          & $1.5 \times 10^{-2}$ & $2.6 \times 10^{-2}$ & $8.9 \times 10^{-2}$ & $4.2 \times 10^{-1}$ &&&& \\
A3571 \dotfill          & $1.9 \times 10^{-2}$ & $3.4 \times 10^{-2}$ & $1.1 \times 10^{-1}$ & $5.3 \times 10^{-1}$ &&&& \\
                                 
\hline 
\end{tabular}
\end{center}
\label{tab: Cangaroo}
\end{table*}

Based on the previous results we discuss the detectability of IC emission by
secondary CRe and pion decay induced $\gamma$-ray emission by current and
future satellite missions as well as operating and future \v{C}erenkov
telescopes.

\subsection{Detectability of pion decay induced $\gamma$-ray and IC emission by
  secondary CRe with INTEGRAL} 

The imager IBIS which is the Imager on Board the 
``INTErnational Gamma-Ray Astrophysics Laboratory'' 
(INTEGRAL)\footnote{\tt http://astro.esa.int/Integral/} 
Satellite covers an energy range from 15~keV up to 10~MeV and is capable of high
resolution imaging ($12 \arcmin$ FWHM) and source identification.
Its spectral sensitivity reaches down to  
$5\times 10^{-8} ~\gamma \mbox{ s}^{-1} \mbox{ cm}^{-2} 
\mbox{ keV}^{-1}$ ($3\,\sigma$ in $10^6$ s, $\Delta E = E/2$) 
to the continuum at 10~MeV.
However, this is most probably not sufficient in order to detect the pion decay
induced $\gamma$-rays of a particular cluster (compare Fig.~\ref{fig:qgamma}).
Assuming a CRp spectral index of $\alpha_\p = 2.3$ and taking the results of
Table~\ref{tab: Xcrp radio} we expect an IC emission
of hadronically originating CRe in the Perseus cluster of 
\begin{equation}
\frac{\dd \mathcal{F}}{\dd E} (20 \mbox{ keV})=
\mathcal{F}_\mathrm{IC}\, 10^{-7} \,\gamma 
\mbox{ cm}^{-2} \mbox{ s}^{-1} \mbox{ keV}^{-1}, 
\end{equation}
with $\mathcal{F}_\mathrm{IC} = 8.4,$ 4.2, and 2.3 for $B_0 = 5~\mu \mbox{G},
10~\mu \mbox{G},$ and $20~\mu \mbox{G}$.  Comparing these results to the {\em
  post-launch} spectral sensitivity of $4\times 10^{-6} ~\gamma \mbox{ s}^{-1}
\mbox{ cm}^{-2} \mbox{ keV}^{-1}$ to the continuum at 20~keV for an observation
time of $10^6$~s ($3\,\sigma$ detection) there is only a minor chance to detect
IC emission of CRe.  However, for steeper spectral indices or a strongly
inhomogeneously magnetized environment, IC fluxes can be enhanced at the
expense of synchrotron emission according to \citet{1999AA...344..409E}.
 
\subsection{Possibility of pion decay induced $\gamma$-ray detection by GLAST} 

The ``Large Area Telescope'' (LAT) onboard the ``Gamma-ray Large Area Space
Telescope''
(GLAST)\footnote{\tt http://glast.gsfc.nasa.gov/science/} 
scheduled to be launched in 2006 has an angular resolution smaller than
$3.5\degr$ at 100~MeV while covering an energy range of 20~MeV up to 300~GeV
with an energy resolution smaller than 10\%. 
Assuming a photon spectral index of $\alpha_\gamma=2$ for the $\gamma$-ray
background the point source sensitivity at high galactic latitude in an one
year all-sky survey is better than  
$6\times 10^{-9} \mbox{ cm}^{-2} \mbox{ s}^{-1}$ for energies integrated 
above 100~MeV.
Assuming the radio-mini halo in the Perseus cluster mainly to
originate from secondary electrons emitting synchrotron radiation then
we expect the CRp scaling parameter to be typically one order of
magnitude below the upper limits obtained by comparing to EGRET data. 
This immediately would imply a good possibility to detect pion decay induced
$\gamma$-ray emission by GLAST preferentially in nearby cooling flow clusters
like Perseus and Virgo.
Specifically for our secondary model of the radio mini-halo of Perseus, 
while assuming $\alpha_\p=2.3$ in the CRp isobaric model we
expect an integrated $\gamma$-ray flux above 100~MeV from Perseus of 
$\mathcal{F}_\gamma(>100 \mbox{ MeV}) /(\gamma \mbox{ cm}^{-2} \mbox{ s}^{-1})=
1.3 \times 10^{-8}, 7.4 \times 10^{-9},\mbox{ and } 5.6 \times 10^{-9}$ for 
$B_0  = 5\mu \mbox{G}, 10\mu \mbox{G}, \mbox{ and } 20\mu \mbox{G}$. 
The expected $\gamma$-ray flux is ever higher when including lower
energetic photons.

\subsection{Expected $\gamma$-ray flux for \v{C}erenkov telescopes}

In the near future there will be different \v{C}erenkov telescope experiments
operating with several telescopes simultaneously and
therefore allowing stereoscopic observations.
On the southern hemisphere there are the ``Collaboration between Australia and
Nippon for a Gamma Ray Observatory in the Outback'' 
(CANGAROO)\footnote{\tt http://www.physics.adelaide.edu.au/astrophysics/} 
in Australia and the ``High Energy Stereoscopic System'' 
(HESS)\footnote{\tt http://www.mpi-hd.mpg.de/hfm/HESS/HESS.html} in Namibia.
On the northern hemisphere there will be the 
``Very Energetic Radiation Imaging Telescope Array System'' 
(VERITAS)\footnote{\tt http://veritas.sao.arizona.edu/} in Arizona.
All these telescopes have comparable lower energy thresholds of $E_\mathrm{thr} =
100$~GeV and provide flux sensitivities better than 
$\mathcal{F}_{\gamma,\,\mathrm{exp}}(E>100 \mbox{ GeV}) = 
10^{-12}\,\gamma \mbox{ cm}^{-2}\mbox{ s}^{-1}$.
On the northern hemisphere there will also be the
``Major Atmospheric Gamma-ray Imaging \v{C}erenkov detector'' 
(MAGIC)\footnote{\tt http://hegra1.mppmu.mpg.de/MAGICWeb/} on the Canary Islands
observing with a single dish telescope of 234~$\mbox{m}^2$ providing an even
lower energy threshold of $E_\mathrm{min} = 30$~GeV.

Following the formalism described in Sect.~\ref{sec: simulated gamma} and
comparing the resulting $\gamma$-ray flux
$\mathcal{F}_\gamma(E>E_\mathrm{thr})$ to expected flux sensitivities of
\v{C}erenkov telescopes
$\mathcal{F}_{\gamma,\,\mathrm{exp}}(E>E_\mathrm{thr})$, we obtain possible
upper limits on the CRp scaling parameter $X_\crp$ for an integrated volume
  out to a radial distance of $3\,h_{70}^{-1}$~Mpc.  Table~\ref{tab: Cangaroo}
shows constraints for $X_\crp$ using the isobaric and the adiabatic model of
CRp described in Sect.~\ref{sec: CRp models}.  By comparing these limits to
those obtained by analyzing synchrotron emission in the Perseus and Coma
cluster (see Table~\ref{tab: Xcrp radio}) and assuming a substantial
contribution of hadronically originating CRe to these radio halos there is a
realistic chance to detect extragalactic pion decay induced $\gamma$-ray
emission in clusters like Perseus (A~426), Virgo, Ophiuchus, and Coma (A~1656).

\section{Conclusion}

We investigated hadronic CRp-p interactions in the ICM of clusters and
simulated the resulting emission mechanisms in radio, X-rays, and $\gamma$-rays
assuming spherical symmetry. By applying this technique to a sample of prominent
clusters of galaxies including cooling flow clusters we succeeded in constraining
the population of CRp. Especially cooling flow regions are perfectly suited for
constraining non-thermal ICM components due to their high gas density and
magnetic field strength.

For the first time we developed an analytic formalism to describe the
$\pi^0$-decay induced $\gamma$-ray spectrum self-consistently for a given
differential number density distribution of the CRp population being described
by a power-law in momentum $p_\p$ and parametrized by the spectral index
$\alpha_\p$.  Assuming a constant scaling between kinetic CRp energy density
  and thermal energy density of the ICM we derived an analytic
$\mathcal{F}_{\gamma}$--$F_\mathrm{X}$ scaling relation which only applies
accurately for isothermal clusters. Given the bolometric X-ray luminosity of a
particular cluster this formula estimates the expected $\gamma$-ray flux
$\mathcal{F}_{\gamma}$ owing to inelastic cosmic ray ion collisions.  From the
literature we collected electron density and temperature profiles of seven
cooling flow clusters and four non-cooling flow clusters using the
$\mathcal{F}_{\gamma}$--$F_\mathrm{X}$ scaling relation to obtain
observationally promising candidates. We furthermore present formulae
describing the synchrotron and inverse Compton emission of hadronically
originating secondary electrons assuming an isotropic distribution of magnetic
fields following a smooth profile.

In order to apply this method to our sample of clusters of galaxies we
introduced three specific models for the spatial distribution of CRp within
cooling flow cluster.  In our first two scenarios we characterized the kinetic
CRp energy density $\eps_{\crp} (\vec{r})$ to be a constant fraction of the
thermal energy density $\eps_\mathrm{th} (\vec{r})$ of the ICM parametrized by
$X_\crp$.  The CRp isobaric model assumes the average pressure of CRp not to
change during the formation of the cooling flow while the adiabatic model
hypothesizes this proportionality prior to transition because the CRp
experience adiabatic compression during the relaxation phase.  In our third
scenario we modeled the resulting distribution of CRp diffusion from a central
source.  By modeling the particular $\gamma$-ray emission of our cluster sample
and comparing to EGRET upper limits we obtained upper bounds on the CRp scaling
parameter $X_\crp=\eps_{\crp} (\vec{r})/\eps_\mathrm{th} (\vec{r})$.  For
Perseus and Virgo we infer the strongest upper limits which lie in the range
$X_\crp\in [0.08,0.18]$ for different choices of the CRp spectral index
$\alpha_\p \in [2.1,2.7]$.

Furthermore, the radio emission due to hadronically produced secondary
electrons emitting synchrotron radiation was calculated and resulting radio
brightness profiles were compared to measured data of the radio-mini halo of
Perseus as well as the radio halo of Coma.  In the case of Coma our CRp
  profiles characterized by a flat CRp scaling parameter $X_\crp$ are not able
  to reproduce the observed radio profiles particularly in the peripheral
  regions of the cluster. In the following we adjusted the radial behavior of
  $X_\crp(r)$ such that the synchrotron emission resulting from hadronic CRe is
  able to account for the observed radio surface brightness profile and thus
  allowing for an additional degree of freedom. The resulting increase of
  $X_\crp(r)$ for larger radii could be due to adiabatic compression which
  increases the thermal energy density at a higher rate than the CRp energy
  density. Even more important, the aspherical Coma cluster morphology reduces
  the required radial increase in $X_\crp(r)$.  By exploring the accessible
  parameter space spanned by parameters describing the magnetic field and the
  spectral index of the CRp population we identify regions where the hadronic
  scenario is able to reproduce the observed radio profiles preferentially for
  an energy density of the magnetic field which declines shallower than the
  thermal energy density.  We conclude that the secondary model for radio halos
  is still viable.

In the case of the Perseus mini-radio halo, we conclude upper limits on
$X_\crp$ which are ranging for the isobaric model of CRp within the interval
$X_\crp \in [0.01,0.1]$ for conservative combinations of values of the magnetic
field $B$ and the CRp spectral index $\alpha_\p$ while upper limits for the CRp
adiabatic model are typically half an order of magnitude below.  By comparing
calculated radio brightness profiles to measured data of the radio-mini halo in
Perseus, we found excellent morphological agreement between the CRp isobaric
model and the radio data especially for the choice of $B_0 = 10 ~\mu \mbox{G}$,
$\alpha_B = 0.5$, and $\alpha_\p = 2.3$.  In the course of this paper we argued
that this specific choice of parameters for the magnetic fields in cooling flow
clusters is also preferred by experiments like Faraday rotation measurements
and cosmological cluster simulations including magnetic fields.  A
discussion of different acceleration mechanisms of CRp such as structure
formation shocks, supernovae remnants, and injection by active radio galaxies
supports also a value of $\alpha_\p$ close to the inferred one.  Because of
the required moderate CRp energy density we propose synchrotron radiation by
non-thermal secondary electrons from hadronic interactions as a likely
explanation of radio mini-halos.  In order to scrutinize this model we provide
predictions of $\gamma$-ray fluxes for \v{C}erenkov telescopes as well as the
INTEGRAL and GLAST satellites.

Finally, we analyzed the possibility of detecting such pion decay induced
$\gamma$-ray and IC emission by current and future satellite missions as well
as new generation \v{C}erenkov telescopes.
Depending on the CRp spectral index, the fragmentation of the spatial
distribution of the magnetic field as well as its field strength, it will be
difficult for INTEGRAL to detect the IC emission of the hadronically
originating secondary CRe while GLAST has the potentiality to detect the
distinct signature of the pion decay induced $\gamma$-ray emission
preferentially in nearby cooling flow clusters.
By investigating the opportunity of detecting extragalactic $\gamma$-rays
by \v{C}erenkov telescopes we argued in favor of four candidate clusters
(Perseus (A~426), Virgo, Ophiuchus, and Coma (A~1656)) which are
especially suited to detect hadronically originating $\gamma$-ray emission.

\begin{acknowledgements}
  In particular we are indebted to Francesco Miniati for fruitful discussions
  and providing numerical $\gamma$-ray spectra. We also wish to thank Matthias
  Bartelmann, Bj\"orn Malte Sch\"afer and an anonymous referee for carefully
  reading the manuscript and their numerous constructive remarks. Furthermore,
  we acknowledge useful discussions with Eugene Churazov and Sebastian Heinz.
  This work was performed within the framework of the European Community
  Research and Training Network {\it The Physics of the Intergalactic Medium}.
\end{acknowledgements}

\appendix

\section{Deprojection of X-ray surface brightness profiles represented by double
  $\beta$-models} 
\label{sec: deprojection}

Owing to the enhanced electron density in the central region the X-ray surface
brightness profile $S_\mathrm{X} (r_\bot)$ in cooling flow cluster can be
represented by double $\beta$ models,
\begin{equation}
S_\mathrm{X} (r_\bot)= \sum_{i=1}^2 \,S_i\, 
\left[ 1 + \left( \frac{r_\bot}{r_{\mathrm{c}_i}}\right)^2\right]
^{-3\beta_i + 1/2},
\label{double beta}
\end{equation}
where the X-ray surface brightness profile is a line of sight projection of the
squared electron density and the cooling function relative to the squared
electron density $\tilde{\Lambda}_\mathrm{X}(T_\e)$,
\begin{eqnarray}
S_\mathrm{X} (r_\bot) &=& 
\int_{-\infty}^\infty \dd z \,n_\e^2\left( \sqrt{r_\bot^2+ z^2}\right)
\,\tilde{\Lambda}_\mathrm{X} \left[ T_\e \left( \sqrt{r_\bot^2+ z^2}\right)\right] \\
&=& 2 \int_{r_\bot}^\infty \dd r \,\frac{r\,n_\e^2 (r) \, 
\tilde{\Lambda}_\mathrm{X}[T_\e(r)]}{\sqrt{r^2 - r_\bot^2}}\,.
\end{eqnarray}
Thus the electron density $n_\e(r)$ can be derived from $S_\mathrm{X} (r_\bot)$
by inverting the Abel equation
\begin{eqnarray}
n_\e^2 (r) \, \tilde{\Lambda}_\mathrm{X}[T_\e(r)] &=& -\frac{1}{\pi\,r}\,\frac{\dd}{\dd r}
\int_r^\infty\dd y \,\frac{y \,S_\mathrm{X}(y)}{\sqrt{y^2 - r^2}}\\
&=& -\frac{1}{\pi}\int_r^\infty\dd y \,
\frac{S_\mathrm{X}'(y)}{\sqrt{y^2 - r^2}}\,,
\end{eqnarray}
where the prime denotes the derivative. For the second equation we used that
$n_\e(r)$ is bounded for $r \to \infty$.
Using eq.~(\ref{double beta}) this equation can be solved analytically yielding
\begin{equation}
n_\e^2 (r) = \,\frac{1}{\tilde{\Lambda}_\mathrm{X}[T_\e(r)]}
\sum_{i=1}^2 \frac{S_i}{2\pi\,r_{\mathrm{c}_i}}\,
    \frac{6 \beta_i - 1}{\left(1 + r^2/r^2_{\mathrm{c}_i}\right)^{3\beta_i}}\,
    \mathcal{B}\left(\frac{1}{2},3\beta_i\right)\,,
\end{equation}
where $\mathcal{B}(a,b)$ denotes the beta-function \citep{1965hmfw.book.....A}.
Provided the central density $n_\e(0)$ is known and assuming furthermore the
special case of equality of the two $\beta$ parameter, $\beta_1 = \beta_2$, we
arrive at the following compact formula for the electron density profile $n_\e
(r)$
\begin{eqnarray}
n_\e (r) &=& \,\left[\frac{\tilde{\Lambda}_\mathrm{X}[T_\e(0)]}{\tilde{\Lambda}_\mathrm{X}[T_\e(r)]} 
\times\sum_{i=1}^2 n_i^2 \, \left( 1+\frac{r^2}{r^2_{\mathrm{c}_i}}\right)^{-3\,\beta}
\right]^{1/2}\!\!,\\
n_i &=& n_\e(0)\,\left(\sum_{j=1}^2\frac{S_j\,r_{\mathrm{c}_i}}{S_i\,r_{\mathrm{c}_j}}
\right)^{-1/2}\!\!.
\end{eqnarray}
Generalizing to n-fold $\beta$-profiles can be obtained by means of induction.

\bibliography{bibtex/chp}
\bibliographystyle{aa}

\end{document}